\documentclass[twocolumn,showpacs,preprintnumbers,amsmath,amssymb,superscriptaddress]{revtex4}
\usepackage{graphicx}
\usepackage{dcolumn}
\usepackage{bm}
\usepackage{color}

\newcommand{\ve}[1][K]{\mathbf{#1}}

\def \equi#1{\mathrel{\mathop{\kern 0pt\sim}\limits_{#1}}} 

\begin{document}

\title{Universal first-passage  statistics of aging processes}

\author{N. Levernier}
 \affiliation{Laboratoire de Physique Th\'eorique de la Mati\`ere Condens\'ee, CNRS/UPMC, 
 4 Place Jussieu, 75005 Paris, France}
 
 \author{O. B\'enichou}
\affiliation{Laboratoire de Physique Th\'eorique de la Mati\`ere Condens\'ee, CNRS/UPMC, 
 4 Place Jussieu, 75005 Paris, France}
 
 \author{T. Gu\'erin}
 \affiliation{Laboratoire Ondes et Mati\`ere d'Aquitaine, University of Bordeaux, Unit\'e Mixte de Recherche 5798, CNRS, F-33400 Talence, France }
 
 \author{R. Voituriez}
\affiliation{Laboratoire Jean Perrin, CNRS/UPMC, 
 4 Place Jussieu, 75005 Paris, France}

\begin{abstract} Many out of equilibrium phenomena, such as diffusion-limited reactions  or target search processes,
 are controlled by first-passage events.  
So far the general determination of the mean first-passage time (FPT) to a target in confinement  has left aside aging processes, involved in contexts as varied as glassy dynamics, tracer diffusion in
biological membranes or transport of cold atoms in optical lattices.  Here we consider general non-Markovian scale-invariant processes in arbitrary dimension, displaying  aging,  and demonstrate that all the moments of the FPT obey universal scalings with the confining volume with non trivial exponents. Our analysis shows that a nonlinear scaling of the mean FPT with the volume  is the hallmark of aging and provides a general tool to quantify its impact  on first-passage kinetics in confinement.
\end{abstract}

\maketitle

 How long does it take  a random walker to find a target site ? 
This time, usually  called 
a first-passage time (FPT),  has been extensively studied  \cite{Redner:2001a,Metzler:2014,Benichou:2014} because of its relevance to the many physical processes that  are controlled 
by first-passage events.  
At the microscopic scale, this is the case of diffusion-limited reactions, which are controlled by the encounter of  reaction partners \cite{Rice:1985}.  At the macroscopic scale, random search processes, exemplified by the search of food by  animals, have been shown over the last decade to be efficiently quantified by  FPTs \cite{Viswanathan:1999a,Edwards:2007,Shlesinger:2006,Condamin:2007zl,Guerin:2016}.
 A first step in the analysis of FPTs consists in determining its mean, the mean first-passage time (MFPT).

Geometrical confinement  is  a key parameter in the evaluation of MFPTs, as  illustrated on the example of a $1d$ or $2d$  symmetric random walk with nearest-neighbors jumps on a regular lattice.   In the absence of  confinement  it is well known that a $1d$ or $2d$ random walker  eventually visits any site of the lattice with certainty. However, due to the long tail statistics of the FPT distribution, the MFPT turns out to be infinite in this case. In higher dimensions, the MFPT to a given site is still infinite, since the weight of trajectories that never reach it is finite.  The analysis of  FPT statistics for general random walks in such infinite geometries has been the focus of an intense activity \cite{Krapivsky:1996,Majumdar:1996,Krug:1997,Majumdar:1999a,Molchan1999,Bray:2013} .
In the opposite case where the random walker is confined in a bounded domain, the situation is radically  different.  A target site is found with probability one and the long tails of the FPT distribution are  generally suppressed, leading to  a finite MFPT. The question that naturally arises is then to determine the scaling of the MFPT with the volume of the confining domain. 

\begin{figure}[h!]
\begin{center}
\includegraphics[width=8cm]{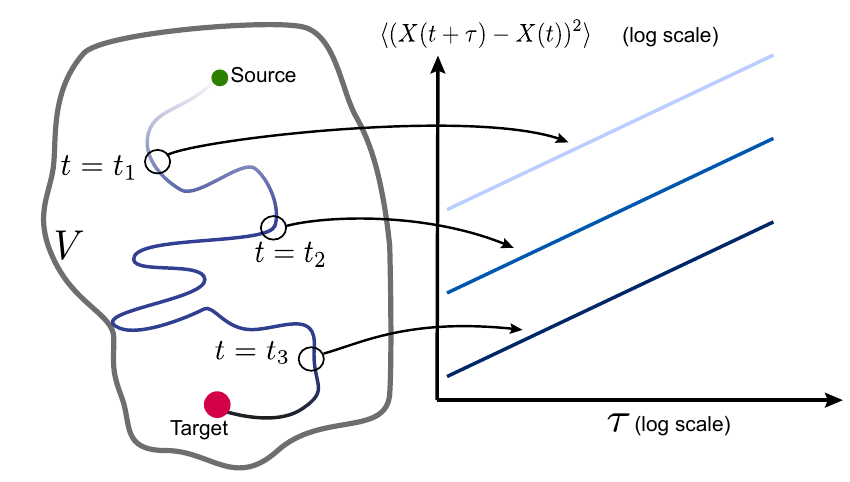}
\end{center}
\caption{\textbf{First-passage times (FPTs) of aging processes.} An aging random walker  is searching for a target in a confining domain of volume $V$, with reflecting boundaries. Aging is characterized by a mean square displacement that depends on the observation time $t$. In this article, we show that the FPT  obeys universal statistics.}
\label{fig1}
\end{figure}

This cannot be determined solely by dimensional analysis since {\it a priori} at least three length scales are involved: the target size $a$, the source to target distance $r$ and the typical confining domain size $R\propto V^{1/d_f}$, where $V$ is the volume of the confining domain and $d_f$ the fractal dimension of the system. 
However, in the case of scale-invariant Markovian processes, a general  scaling of the MFPT with $a$, $r$ and $V$ has been derived in the large volume limit $V\to\infty$, yielding in particular a general linear dependence on $V$ \cite{Condamin:2007zl}. This result  was further extended to higher order moments and the full FPT distribution was determined asymptotically \cite{BenichouO.:2010a}.  
Recently, the case of non-Markovian Gaussian processes with stationary increments was analyzed \cite{GuECrinT.:2012,Guerin:2016} and a linear scaling with $V$ was again found for the MFPT. However, these analyses  leave aside the class of general aging processes \footnote{Note that  continuous time random walks  (CTRWs)  display aging \cite{Schulz:2014}. However,  as shown in SI, this aging has no effect on the FPT statistics in the limit of large confining volume considered here (see also \cite{Condamin:2007yg,Condamin:2008,Krusemann:2014}).},  
 for which physical observables depend on the time elapsed since the preparation of the system. Physical examples of aging processes range from the  slowing down  observed in glassy systems \cite{Berthier:2011} to the broadening in the velocity distribution of cold atoms in optical lattices \cite{Dechant:2012} or the non-stationary behavior of the mean square displacement of tracers in the plasma membrane \cite{Weigel:2011}.
More generally, any random walker interacting with other degrees of freedom is expected to display memory effects, in which case aging properties, reflecting the decay of the memory of the initial state, naturally arise.  Here, we show that  in the general case of scale-invariant non-Markovian processes in confinement, displaying potentially aging properties, 
the full FPT distribution  falls into universality classes that we define below.




We consider a general scale-invariant non-Markovian stochastic process defined in a  space of (potentially fractal) dimension $d_f$, and characterized by its walk dimension $d_w$ defined through 
$\langle X^2(t)\rangle \equi{t\to\infty} t^{2/d_w}$ and symmetric increments satisfying at long times:  $\langle (X(t+\tau)-X(t))^2\rangle \equi{t\to\infty} t^{\alpha} \tau^{2/d_w-\alpha}$. Here, we introduce an aging exponent $\alpha$  which describes for $\alpha\neq0$ the non stationary dynamics of increments ; $\alpha>0$ corresponds qualitatively to accelerating processes and $\alpha<0$ to slowing down processes. Note that this class of processes covers a wide range of examples as detailed below. It is useful to distinguish between compact  and non-compact processes that we define here by the behavior of the probability $P(a,r)=P \left(\frac{a}{r}\right)$ that in unbounded space  a target of size $a$ is eventually reached by the  process that started from a distance $r$. (i) If $P \left(\frac{a}{r}\right)=1$ for all $a$, the process is called compact and  even a point like target is found with certainty ($P(0)=1$). One then introduces the survival probability  $S$ that a point like target has not been reached up to time $t$, which decays as $S(t)\equi{t\to\infty} t^{-\theta}$ where the  persistence exponent $\theta$ has been extensively studied in the literature \cite{Krapivsky:1996,Majumdar:1996,Krug:1997,Majumdar:1999a,Bray:2013}. (ii) Conversely, if $P(0)=0$, the process is termed non-compact, and will be characterized by the transience exponent $\psi$ that we introduce here and define by the small $a/r$ scaling of $P$ : $P \left(\frac{a}{r}\right)\equi{a\to0} \left( \frac{a}{r}\right)^\psi$.  (iii) Finally, the case of marginal exploration is defined here by
$S(t)\sim1/\ln t$ for a target of radius $a\neq0$.

We report here  that the distribution of the FPT to a target in a confined domain of volume $V\sim R^{d_f}$ (see Fig.\ref{fig1}) obeys  universal  statistics defined as follows. Denoting by $T$ the FPT, we introduce the  rescaled random variable $\eta\equiv T/T_{\rm typ}$, where the characteristic time  $T_{\rm typ}$ is defined by:
\begin{equation}
\label{Ttyp}
T_{\rm typ} =  \left \{
\begin{array}{ll}
R^{d_w} & \mbox{(compact)} \\
R^{d_w} \left(\ln \frac{R}{a}\right)^{1-\alpha d_w/2} & \mbox{(marginally compact)} \\
R^{d_w} \left( \frac{R}{a}\right)^{\psi(1-\alpha d_w/2)} & \mbox{(non-compact)}.
\end{array}
\right .
\end{equation}
Assuming that the mean FPT to the target  is finite (processes leading to infinite MFPTs, such as CTRWs with infinite mean waiting times, are analyzed in SI), we  find that $\eta$ is asymptotically distributed in the large $R$ limit with the distribution:
\begin{equation}
\label{main}
G(\eta;a,r,R)=   \left \{
\begin{array}{ll}
h(\eta)\left(\frac{r}{R}\right)^{d_w \theta}& \mbox{(compact)} \\ 
h(\eta) \frac{\ln\frac{r}{a}}{\ln\frac{R}{a}} & \mbox{(marginally compact)} \\
h(\eta)\left[1-C\left(\frac{a}{r}\right)^\psi \right]& \mbox{(non-compact)} \\
\end{array}
\right .
\end{equation}
where  it is assumed that $a\ll r$, $h(\eta)$ is an undetermined scaling function {\it a priori} process-dependent and $C$ is a numerical constant.   In addition, it is argued in SI that 
\begin{equation}
\label{psi}
\psi=d_f-\frac{d_w}{1-\alpha d_w/2}.
\end{equation}

This explicitly determines the dependence on the geometrical parameters of the problem of the FPT distribution, and therefore of all its moments (when they exist). In particular, the scaling of the mean is given by:
\begin{equation}
\label{Tmin}
\langle T\rangle\sim   \left \{
\begin{array}{ll}
R^{d_w(1-\theta)}r{^{d_w \theta}}& \mbox{(compact)} \\
 \frac{R^{d_w}}{\left(\ln R / a\right)^{\alpha d_w/2}} \ln\frac{r}{a}& \mbox{(marginally compact)} \\
\frac{R^{d_w+\psi(1-\alpha d_w/2)}}{a^{\psi(1-\alpha d_w/2)}} \left[1-C\left(\frac{a}{r}\right)^\psi \right]& \mbox{(non-compact)} \\
\end{array}
\right ..
\end{equation}
Remarkably, the scaling of the MFPT can therefore be non  linear with the volume $V\sim R^{d_f}$ for non-trivial values 
of the persistence exponent $\theta$ (in the compact case) or of the aging exponent $\alpha$ (for non-compact processes). This will be the case of generic processes with non stationary increments, as   discussed below.

\begin{figure}[h!]
\begin{center}
\includegraphics[width=9cm]{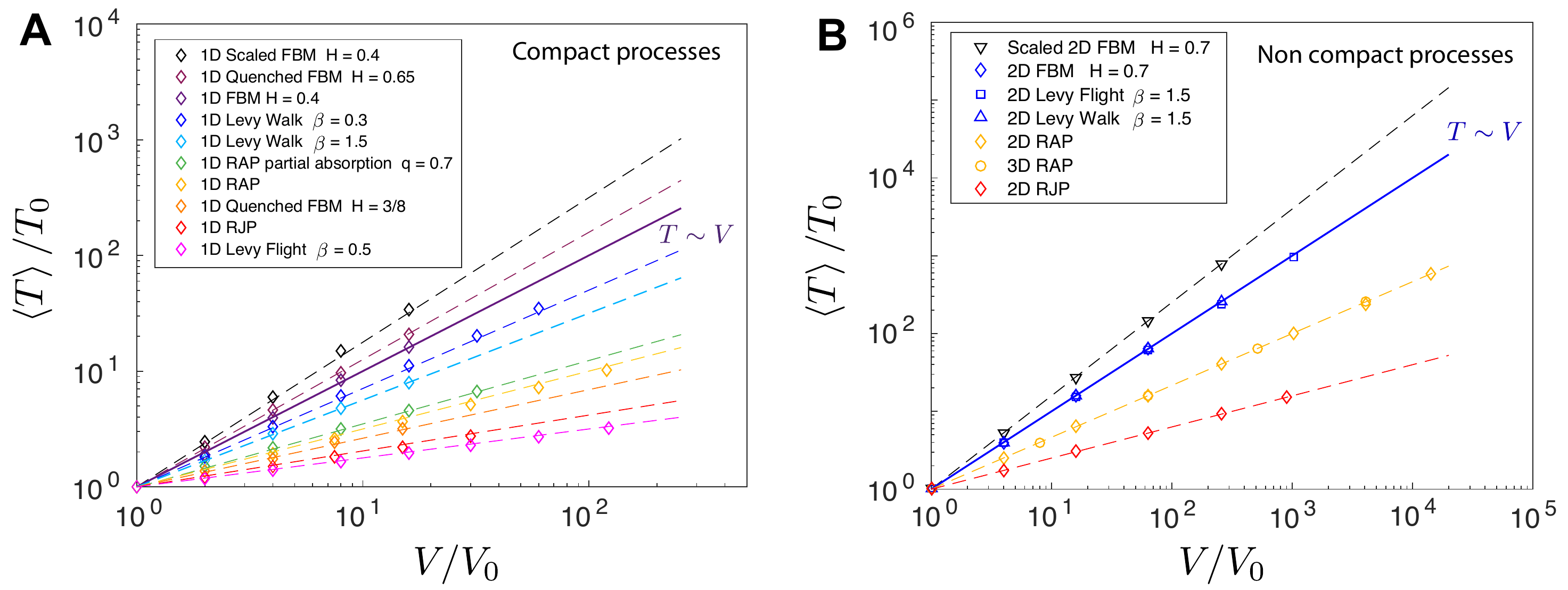}
\end{center}
\caption{\textbf{Mean FPT as a function of the volume for compact (A) and non-compact processes (B).  } Lines represent our  predictions [Eq.\eqref{Tmin}], symbols are the results of numerical simulations. For each process, $\langle T\rangle$ and $V$ are rescaled by arbitrary factors $(V_{0},T_{0})$; $V_{0}$ is chosen big enough to have reached the large volume limit. The continuous  line corresponds to the linear behavior with the volume.  
Details on simulation algorithms and geometrical parameters for each process are given in SI. 
}
\label{fig2}
\end{figure}

We now sketch the demonstration of this result (see SI for details) and first consider the compact case, for which $a$ can be taken equal to zero as stated above. We write the FPT distribution $F(t)$ as a partition over trajectories reaching either the reflecting boundary before the target (with probability $\pi(r,R)$ and conditional FPT distribution $F_b$) or the target first (with probability $1-\pi(r,R)$ and conditional FPT distribution $F_t$): $F=\pi F_b + (1-\pi) F_t$. We now evaluate each of the terms of this equation. We remark that $1-\pi$ can be written as the time integral of the FPT density to the target restricted to trajectories reaching the target before the boundary. Most of these events occur within the typical time scale $R^{d_w}$ needed to reach the boundary, so that one can write
\begin{equation}
1-\pi(r,R)\equi{R\gg r}\int_0^{A R^{d_w}}F_\infty(t,r){\rm d}t,
\end{equation}
 where $F_\infty(t,r)$ is the FPT density in unconfined space and $A$ is a constant. 
Using the scale-invariance of the process, which gives $F_\infty(t,r)\equiv f(t/r^{d_w})/t$, together with the definition of $\theta$, we obtain 
 \begin{equation}
\pi(r,R)\equi{R\gg r}\left(\frac{r}{R}\right)^{d_w \theta}.
\label{pii}
\end{equation}
Note that this  extends the one dimensional result of Majumdar et al. \cite{Majumdar:2010}. The above argument also implies that
\begin{equation}
F_t(t,r,R)\sim Y(AR^{d_w}-t) F_\infty(t,r)\sim Y\left(A-\frac{t}{R^{d_w}}\right)\frac{r^{d_w \theta}}{t^{\theta+1}}
\end{equation}
where $Y$ stands for the Heaviside step function. 
Making use again of scale-invariance and writing
\begin{equation}
F_b(t,R)\equiv g(t/R^{d_w})/t,
\end{equation}
we finally obtain Eq(\ref{main}) above. 

We now turn to the non-compact case. We define by excursion a portion of trajectory starting from the sphere $S$ of radius $R/2$ centered on the target, reaching the boundary and coming back to $S$. We write the FPT distribution as a partition over the number $n$ of excursions  before the first-passage to the target and denote by $\Phi_n(t)$ the corresponding conditional FPT distribution : 
\begin{equation}
F(t,a,r,R)=p_0 \Phi_0(t) + \sum_{n=1}^\infty \Phi_n(t)(1-p_0)(1-p)^{n-1}p,
\label{nc1}
\end{equation}
where $p\sim (a/R)^\psi$  is   the probability to reach the target before the boundary starting from $S$  and $p_0\sim (a/r)^\psi$  is the probability to reach the target before the boundary  starting from $r$. Note that here we implicitly assume that excursions are independent in the large $R$ limit. Physically, it originates from  the divergence with $R$ of the typical time $\tau_n$ needed to perform the $n^{th}$ excursion, which hence can be taken larger than  all correlation times of the process,  as was checked numerically (see SI). Finally, 
a  scaling argument shows that
\begin{equation}
\Phi_n(t)=\frac{1}{t}\phi(t/t_n)
\end{equation}
where $t_n$ is the typical time elapsed after $n$ excursions, which satisfies :
\begin{equation}
\langle (X(t_n+\tau_n)-X(t_n))^2\rangle \equi{t_n\to\infty} t_n^{\alpha} \tau_n^{2/d_w-\alpha}=R^2.
\end{equation}
 Using that $t_n=\tau_1+\dots+\tau_n$, this equation leads to 
\begin{equation}
t_n\equi{R\to\infty}R^{d_w}n^{1-\alpha d_w/2}.
\end{equation}
Making use of the definition of $\psi$, the large $R$ behavior of (\ref{nc1}) is found  to lead to Eq(\ref{main}) above. 
\begin{figure}[h!]
\begin{center}
\includegraphics[width=8.5cm]{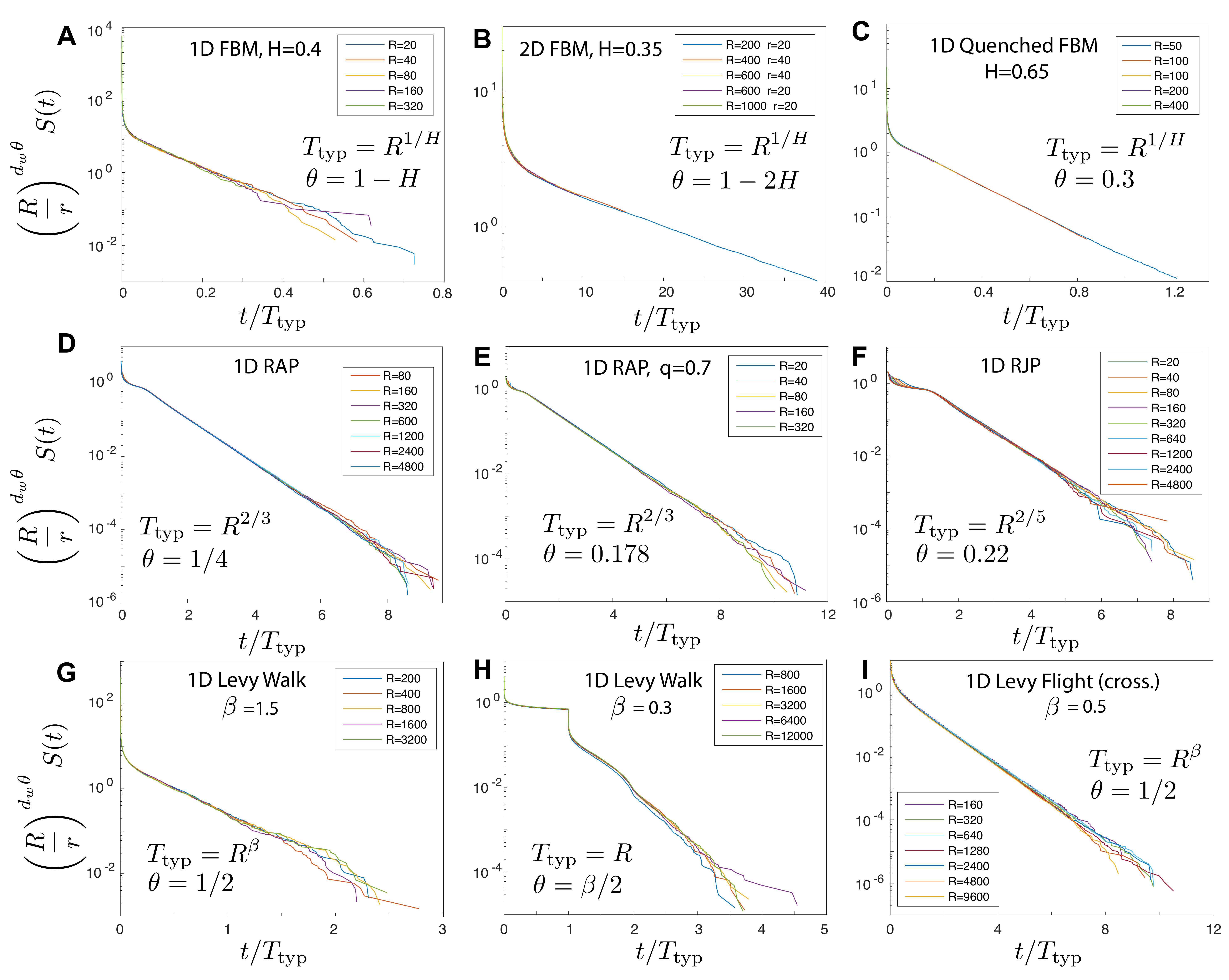}
\end{center}
\caption{\textbf{Universal scalings of the FPT density with geometrical parameters for compact processes.} $S\equiv\int_{t/T_{typ}}^{+\infty}G(\eta	) d\eta$ is the survival probability of the random walker, whose scaling with geometrical parameters is deduced from Eqs.\eqref{Ttyp}, \eqref{main}. The collapse of numerical data after rescaling for different geometrical parameters shows that our theory unambiguously captures the dependence of the FPT distribution on geometrical parameters. Details on the definition of the processes, the simulation algorithms and the geometrical parameters for each process are given in SI. \textbf{A}. $1d$ Fractional Brownian Motion (FBM) with $H=0.4$. \textbf{B}. $2d$ Fractional Brownian Motion with $H=0.35$. \textbf{C}. $1d$ "initially quenched Fractional Brownian Motion", with $H=0.65$. \textbf{D}. $1d$ Random Acceleration Process (RAP). \textbf{E}. $1d$ Random Acceleration Process with a probability $q=0.7$ of absorption at each crossing of the target. \textbf{F}. $1d$ Random Jerk Process (RJP). \textbf{G}. $1d$ L\'evy Walk (XY convention) whose jumps are L\'evy stable distributed with a parameter $\beta=1.5$. \textbf{H}. $1d$ L\'evy Walk (XY convention) whose jumps are L\'evy stable distributed with a parameter $\beta=0.3$. \textbf{I}. $1d$ L\'evy Flight (XY convention) whose jumps are L\'evy stable distributed with a parameter $\beta=0.3$ and the target is found when crossed.}
\label{fig3}
\end{figure}

We now comment on the main results of this paper, Equations \eqref{Ttyp}-\eqref{Tmin}. 
(i) First, the dependence on  $a,r$ and $R$  of the FPT distribution of Markovian processes \cite{BenichouO.:2010a} and the mean FPT of non-Markovian Gaussian processes with stationary increments  \cite{Guerin:2016} are recovered. This results from the specific values of $\alpha$ and $\theta$ for processes with stationary increments:  $\alpha=0$ by definition of stationary increments, which leads to $\psi=d_f-d_w$ for non-compact processes; for compact processes,  we argue in SI that  $\theta=1-d_f/d_w$.
 (ii) Second, a remarkable  feature emerging with aging is  a possible non-linear scaling of the MFPT with the volume, either sub or super linear (see Fig. \ref{fig2}). A linear scaling is shown to hold in the compact case only if increments are stationary at all times (leading to $\theta=1-d_f/d_w$), while in the non-compact case asymptotically  stationary increments are sufficient.  For example, a non-Markovian compact process, that is initially quenched and relaxes to its stationary state in free space, is shown below to display a non-linear scaling of the MFPT with the volume (see Fig. \ref{fig2}).(iii) 
 Third, the FPT statistics displays a strong dependence on the initial distance in the case of compact exploration, and a much weaker one in the non-compact case. Qualitatively, this feature is the same as that previously found for Markovian processes \cite{Condamin:2007zl,BenichouO.:2010a}, but aging quantitatively modifies the exponents characterizing this dependence.
   (iv) Next, note that the time scale $T_{\rm typ}$ of the FPT distribution is independent of aging for compact processes, while it depends explicitly on $\alpha$ for non-compact processes. 
(v) Finally, the  dependence of the FPT distribution on $r,a$ and $R$ falls into universality classes defined by  $d_w$ and $\theta$ (for compact processes) or by   $\alpha,d_w,d_f$ (for non-compact processes).



\begin{figure}[h!]
\begin{center}
\includegraphics[width=8.5cm]{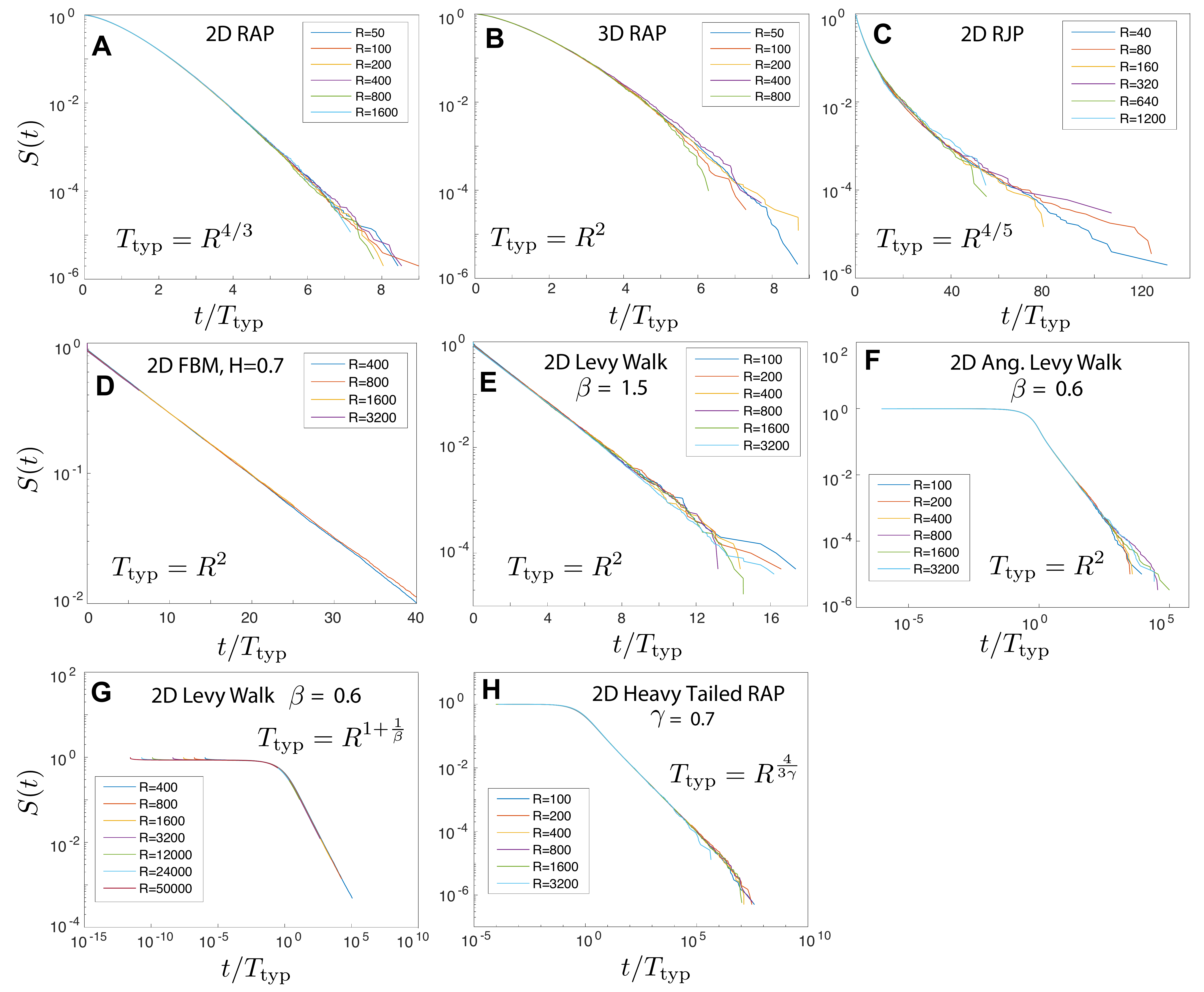}
\end{center}
\caption{\textbf{Universal scalings of the FPT density with geometrical parameters for non-compact processes.} $S\equiv\int_{t/T_{typ}}^{+\infty}G(\eta	) d\eta$ is the survival probability of the random walker, whose scaling on geometrical parameters is deduced from Eqs.\eqref{Ttyp}, \eqref{main}. The collapse of numerical data after rescaling for different geometrical parameters shows that our theory unambiguously captures the dependence of the FPT distribution on geometrical parameters. Details on the processes, the simulation algorithms and the geometrical parameters for each process are found in SI. \textbf{A}. $2d$ Random Acceleration Process. \textbf{B}. $3d$ Random Acceleration Process. \textbf{C}. $2d$ Random Jerk Process. \textbf{D}. $2d$ Fractional Brownian Motion with $H=0.7$. \textbf{E}. $2d$ L\'evy Walk (XY convention) whose jumps are L\'evy stable distributed with a parameter $\beta=1.5$. \textbf{F}. $2d$ L\'evy Walk (angular convention) whose jumps are L\'evy stable distributed with a parameter $\beta=0.6$. \textbf{G}. $2d$ L\'evy Walk (XY convention) whose jumps are L\'evy stable distributed with a parameter $\beta=0.6$. \textbf{H}. So-called $2d$ Heavy Tailed RAP, with a CTRW parameter $\gamma=0.7$. Graphs G and H cover the case of infinite mean FPT, for which Eq.(1) and (3) have to be corrected, as explained in SI. }
\label{fig4}
\end{figure}


We further confirm the validity of our analytical results by comparing them to numerical simulations of a broad range of representative  examples of stochastic processes (see SI for details), for which only sparse results on the FPT statistics in confinement were available so far \cite{PhysRevA.34.2351,Buldyrev:2001a,Burkhardt:2007,Krusemann:2014,Guerin:2016}. Specifically, we consider (1) the $d$-dimensional fractional Brownian motion (FBM), a  non-Markovian Gaussian process of constant mean, with stationary increments satisfying  $\langle [X(t)-X(0)]^2\rangle=t^{2H}$, where $H$ is the Hurst exponent; this process  has been repeatedly invoked in the literature to model anomalous diffusion arising from the interaction with many variables \cite{Krug:1997,Molchan1999};
(2) its extension to quenched initial conditions, for which increments are time dependent and relax asymptotically to the stationary behavior; (3) the $d-$dimensional random acceleration process (RAP), defined by $\ddot{X}=\eta(t)$ with $\eta(t)$ a Gaussian white noise; (4) the generalizations of the RAP to the case of partial absorption,   various conditions at the confining boundary,  higher order derivatives, such as the random jerk process (RJP, satisfying $\dddot{X}=\eta(t)$), and potentially long waiting times;
(5) $d$-dimensional L\'evy flights, where at each time step the random walker performs a jump whose size $l$ is drawn from a long tailed distribution $p(l)\sim 1/l^{1+\beta}$, with both prescriptions of first-arrival  and first-crossing of the target \cite{Chechkin:2003}; (6) $d$-dimensional L\'evy walks, which can be described as  L\'evy flights with a   finite velocity; (7) scaled processes defined from a reference process $X^{(0)}(t)$ by $X(t)\equiv X^{(0)}(t^b)$.  These cases cover  a broad class of stochastic processes in one and higher dimensions, which can be aging or not, Markovian  or non-Markovian, sub or superdiffusive.

Figures \ref{fig2}, \ref{fig3} and  \ref{fig4} reveal excellent quantitative agreement between numerical simulations and our analytical results. The data collapse of the properly rescaled FPT distribution
shows that our approach unambiguously captures the dependence on both $a,r$ and $R$ for both compact (Figure \ref{fig3}) and non-compact (Figure \ref{fig4}) processes.
In particular, sublinear, linear and superlinear  scalings of the mean FPT with the volume are observed, in agreement with our predictions (Figure \ref{fig2}). This demonstrates that the non-linear scaling of the mean FPT with the volume is the hallmark of the aging properties of the dynamics.






\onecolumngrid

\newpage

\begin{center}

\Large{\textbf{Supplemental Material}}

\end{center}

\section{Detailed derivation of the scaling form of the FPT distribution [Eqs.(1)-(4) of the main text]}

\subsection{Detailed derivation of the scaling form of the FPT distribution in compact case}

We derive here in detail the scaling form taken by the FPT distribution in  the compact case. As explained in the main text, in the compact case, the radius $a$ of the target can be taken equal to zero when we focus on the limit $a\ll r$. In this section, we thus have $a=0$. 

Our starting point consists in writing the FPT distribution $F$ as a partition over trajectories reaching either the reflecting boundary 
before the target (with probability $\pi(r,R)$ and conditional FPT distribution $F_b$) or the target first (with probability $1-\pi(r,R)$ and conditional FPT distribution $F_t$): 
\begin{equation}
F(t,r,R)=\pi F_b (t,r,R)+ (1-\pi) F_t(t,r,R).
\end{equation}
We now evaluate each of the terms of this equation. 

We remark that $1-\pi$ can be written as the time integral of the FPT density to the target restricted to trajectories reaching the target before the boundary. Most of these events occur within the typical time scale $R^{d_w}$ needed to reach the boundary, so that one can write
\begin{equation}
1-\pi(r,R)\equi{R\gg r}\int_0^{A R^{d_w}}F_\infty(t,r){\rm d}t,\label{05423}
\end{equation}
where $A$ is a constant and $F_\infty(t,r)$ is the FPT density in unconfined space. Using the scale-invariance of the process, we can write $F_\infty$ under the form
 \begin{equation}
F_\infty(t,r)\equiv \frac{1}{t}f\left(\frac{t}{r^{d_w}}\right),
\end{equation}
where $f$ is a dimensionless function. Using the above equation and the fact that $F_\infty$ is normalized to unity, we obtain from Eq.(\ref{05423})
\begin{equation}
\pi(r,R)\equi{R\gg r}\int_{A R^{d_w}}^\infty \frac{1}{t}f\left(\frac{t}{r^{d_w}}\right){\rm d}t. \label{05U3J}
\end{equation}
Next, using the definition of the persistence exponent, we find that the long time behavior of the function $f$ satisfies
\begin{equation}
f\left(\frac{t}{r^{d_w}}\right)\equi{t\to\infty}\left(\frac{r^{d_w}}{t}\right)^\theta.
\end{equation}
Inserting this relation into the integral (\ref{05U3J}), we finally obtain the following scaling for the probability $\pi$ of touching the boundaries of the confining domain before the target: 
  \begin{equation}
\pi(r,R)\equi{R\gg r}\left(\frac{r}{R}\right)^{d_w \theta}.
\end{equation}
Note that the above relation extends the one dimensional result of \cite{Majumdar:2010}. 

Similarly, we can write the  FPT distribution to the target $F_t$, conditional to the fact that the target is hit before the domain boundaries, as 
\begin{equation}
F_t(t,r,R)\sim Y(AR^{d_w}-t) F_\infty(t,r)\sim Y\left(A-\frac{t}{R^{d_w}}\right)\frac{f(t/r^{d_w})}{t}.
\end{equation}
where $Y$ stands for the Heaviside step function. 

Next, $F_b$ is the conditional FPT distribution for trajectories that hit the domain boundaries first, we can thus argue that it does not depend on $r$. Making use again of scale-invariance, we obtain
\begin{equation}
F_b(t,r,R)\equiv \frac{1}{t}g\left(\frac{t}{R^{d_w}}\right).
\end{equation}
Collecting all terms, we  obtain  
\begin{equation}
F(t,r,R)\equi{R\gg r} \left(\frac{r}{R}\right)^{d_w \theta}\frac{1}{t}g\left(\frac{t}{R^{d_w}}\right) +Y\left(A-\frac{t}{R^{d_w}}\right)\frac{1}{t}f\left(\frac{t}{r^{d_w}}\right). 
\end{equation}
We consider  time scales larger than $r^{d_w}$ and obtain
\begin{equation}
F(t,r,R)\equi{R\gg r} \left(\frac{r}{R}\right)^{d_w \theta}\frac{1}{t}g\left(\frac{t}{R^{d_w}}\right) +Y\left(A-\frac{t}{R^{d_w}}\right)\frac{1}{t}\left(\frac{r^{d_w}}{t}\right)^\theta. \label{095043}
\end{equation}
Using that the distribution $G$ of the rescaled variable $\eta\equiv T/R^{d_w}$ is given by 
\begin{equation}
G(\eta,r,R)=F(t,r,R)R^{d_w},
\end{equation}
we finally obtain that $G$ given by Eq.~(\ref{095043}) can be written as 
\begin{equation}
\label{resultcompact}
G(\eta,r,R)= \left(\frac{r}{R}\right)^{d_w \theta} h(\eta),
\end{equation}
where $h(\eta)=\left(g(\eta)+Y(A-\eta)/\eta^\theta\right)/\eta$ is an undetermined scaling function that is {\it a priori} process dependent. 

\subsection{Detailed derivation of the scaling form of the FPT distribution in the non-compact case}

\label{detailednoncompact}

We derive here in detail the scaling form taken by the FPT distribution in  the non-compact case. Note that we assume here that the mean FPT to the target is finite (in Section \ref{SectionFormalismInfiniteMFPT} we describe how the results are modified in the case of infinite  MFPTs).  

\begin{figure}[h]
\begin{center}
\includegraphics[scale=0.43]{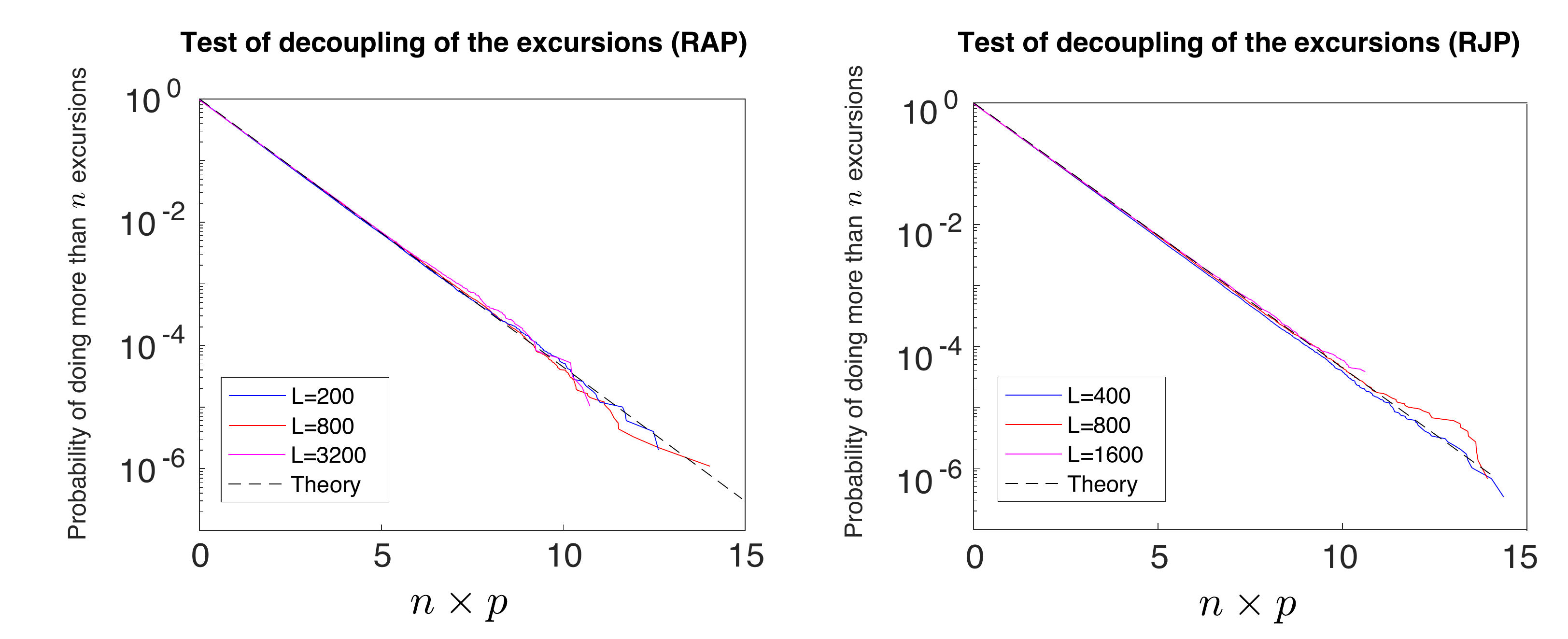} 
\label{FigGeom}
\caption{\textbf{Number of excursions $n$ before reaching the target.} For both processes, the walker begins at position $r=10$ in a 2D domain of linear size $R$, with periodic boundary conditions.  Following our prediction,  $p$ is taken as $p\propto(a/R)$ for both processes. To compare to \eqref{discrete}, we make use of $(1-p)^n\simeq e^{-np}$ for large $R$.}
\end{center}
\end{figure}

We define by excursion a portion of trajectory starting from the sphere $S$ of radius $R/2$ centered on the target, reaching the boundary and coming back to $S$. Let us  denote by $\Phi_n(t)$ the probability density to reach the target at time $t$ given that  $n$ excursions have been performed. 
We write the FPT distribution as a partition over the number $n$ of excursions:
\begin{equation}
\label{discrete}
F(t,a,r,R)=p_0 \Phi_0(t) + \sum_{n=1}^\infty \Phi_n(t)(1-p_0)(1-p)^{n-1}p,
\end{equation}
where $p$  is   the probability to reach the target before the boundary starting from $S$  and $p_0$  is the probability to reach the target before the boundary  starting from $r$. Note that here we implicitly assume that the probability of reaching the target after $n$ excursions is independent of the previous excursions in the large $R$ limit. Physically, it originates from  the divergence with $R$ of the typical time $\tau_n$ needed to perform the $n^{th}$ excursion, which hence can be taken larger than  all correlation times of the process. This geometrical law of the number of excursions before reaching the target was checked numerically for two highly correlated and aging processes : the Random Acceleration and the Random Jerk processes defined below (Fig \ref{FigGeom}).

The probability to find the target during the $n^{th}$ excursion is equal to the probability to find the target of radius $a$ before the domain boundary, starting from a distance $R/2$ from the target. By definition of the  transience exponent $\psi$ introduced in the main text, this probability scales as \begin{equation}
\label{p}
p\equi{R\gg a} C \left(\frac{a}{R}\right)^\psi.
\end{equation}
Similarly, 
\begin{equation}
\label{p0}
p_0\equi{r\gg a} C \left(\frac{a}{r}\right)^\psi.
\end{equation}

We now determine the scalings of the typical time $t_n$ at which the  $n^{th}$ excursion takes place, and the typical time $\tau_n$ between the $n^{th}$ and the $(n+1)^{th}$ excursion. These times can be found by noting that, during $\tau_n$ the typical traveled distance is $R$, which means that
\begin{equation}
\label{deft}
\langle (X(t_n+\tau_n)-X(t_n))^2\rangle \equi{t_n\to\infty} t_n^{\alpha} \tau_n^{2/d_w-\alpha}=R^2.
\end{equation}
The $n$ dependence of $t_n$ can be then found self-consistently. We assume the scaling
\begin{equation}
\label{guesstau}
\tau_n\equi{n\to\infty}\frac{R^{\nu}}{n^\delta},
\end{equation}
where the exponents $\delta$ and $\nu$ will be determined below. Since $t_n=\sum_{k=1}^{n-1}\tau_k$, where the variables $\tau_k$ have finite mean, we obtain
\begin{equation}
t_n\equi{n\to\infty}R^{\nu}n^{1-\delta} 
\end{equation}
The values of $\delta,\nu$ are found by inserting the above expressions into Eq.~\eqref{deft}, leading to
\begin{align}
&\nu=d_w,\\
&\delta=\frac{\alpha d_w}{2}.
\end{align}
Finally,
\begin{align}
&t_n\equi{R\to\infty}R^{d_w}n^{1-\alpha d_w/2}. \label{Expr_tn}\\
&\tau_n\equi{R\to\infty}\frac{R^{d_w}}{n^{\alpha d_w/2}}. \label{Expr_taun}
\end{align}

\begin{figure}[h]
\includegraphics[scale=0.38]{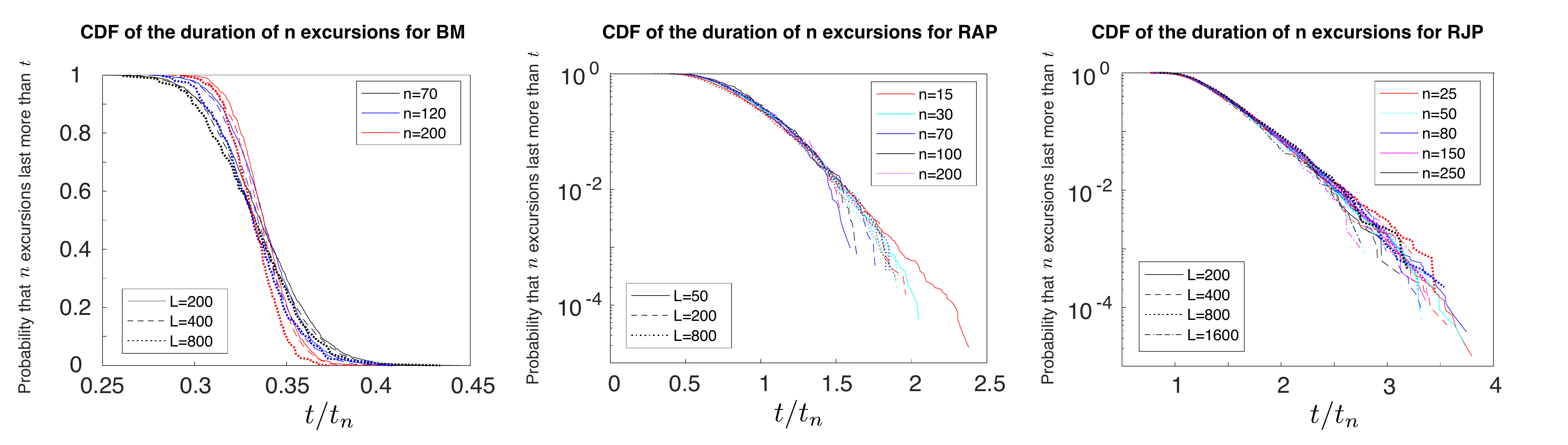} 
\caption{\textbf{Validity of the scaling \ref{scaling test} of the conditional FPT distribution .} We tested the scaling of the conditional FPT distribution with the number of excursions $n$ and the size $R$ of the volume. The first plot presents the result for the regular Brownian motion. By direct application of the Central Limit Theorem, we know that at large $n$ the distribution $\phi_{n}(t)$ is a Gaussian around its mean (which is proportional to $t_{n}\sim n R^{2}$) and variance $\sim \sqrt{n}$. Hence, after rescaling by $t_{n}$, Heaviside step-function is observed. The Random Acceleration Process (RAP) and the Random Jerk Process (RJP) are also considered. In these accelerating processes, the duration of excursions are shorter and shorter and $t_{n}$ is sublinear with $n$. Our predictions for the scaling with $t_{n}=(n R)^{2/3}$ for the RAP and $t_{n}=(n R)^{2/5}$ for the RJP are  verified.}
\label{FigColl}
\end{figure}

Next, we note that $\Phi_n(t)$ can be written as
\begin{equation}
\Phi_n(t)=\frac{1}{t}g\left(\frac{t}{t_n},\frac{R^{d_w}}{t_n},\frac{a}{R}\right).
\end{equation}
Taking the limits $a/R\to0$ and $n\to\infty$ [which implies that $t_n\gg R^{d_w}$ from Eq.~(\ref{Expr_tn})] thus leads to the scaling form:
\begin{equation}
\Phi_n(t)\sim\frac{1}{t}\phi\left(\frac{t}{t_n}\right).
\label{scaling test}
\end{equation}
We check the validity of this scaling on Fig. \ref{FigColl}.

The large $R$ asymptotics of $F(t,a,r,R)$ can be obtained by transforming the discrete sum of Eq.~\eqref{discrete} into an integral
\begin{equation}
F(t,a,r,R)\equi{R\to\infty}p\frac{1-p_0}{1-p}\int_0^\infty \frac{1}{t}\phi\left(\frac{t}{R^{d_w}n^{1-\alpha d_w/2}}\right)e^{n\ln(1-p)}{\rm d}n
\end{equation}
which, by using the change of variables $u=np$ (with $p\ll 1$) and Equations \eqref{p},\eqref{p0}, leads to
\begin{equation}
F(t,a,r,R)\equi{R\to\infty}\left(1-C\left(\frac{a}{r}\right)^\psi\right)\frac{1}{t}\int_0^\infty \phi\left(\frac{t}{R^{d_w}\left[u\left(R/a\right)^\psi/C \right]^{1-\alpha d_w/2}}\right)e^{-u}{\rm d}u,
\end{equation}
where $C$ is a model dependent numerical constant.
A characteristic time appears in the above equation
\begin{align}
T_{\mathrm{typ}}=R^{d_w}\left(R/a\right)^{\psi (1-\alpha d_w/2)}
\end{align}
in terms of which the FPT distribution becomes
\begin{equation}
F(t,a,r,R)\equi{R\to\infty}\left(1-C\left(\frac{a}{r}\right)^\psi\right)\frac{1}{t}\int_0^\infty \phi\left(\frac{t}{T_{\mathrm{typ}}u^{1-\alpha d_w/2}}\right)e^{-u}{\rm d}u.
\end{equation}
We now consider the rescaled time 
\begin{equation}
\eta\equiv \frac{t}{R^{d_w}\left(R/a\right)^{\psi(1-\alpha d_w/2)}}=\frac{t}{T_{\mathrm{typ}} }.
\end{equation}
The distribution $G$ of this rescaled variable is given by 
\begin{equation}
G(\eta,r,R)=F(t,a,r,R)T_{\mathrm{typ}},
\end{equation}
and we finally obtain that $G$ can be written as 
\begin{equation}
\label{U1}
G(\eta,r,R)=  \left(1-C\left(\frac{a}{r}\right)^\psi\right) h(\eta),
\end{equation}
where $h$ is an undetermined scaling function {\it a priori} process dependent, which reads here $h(\eta)=\eta^{-1}\int_0^\infty \phi(\eta u^{(\alpha d_w/2-1)}/C)e^{-u}du$.





Last, in the marginal case, where $\psi=0$, all the steps presented for the non-compact case  hold with the only difference that now :
\begin{equation}
\label{pMarginal}
p\equi{R\gg a}\frac{1}{\ln \left(\frac{R}{a}\right)}
\end{equation}
and
\begin{equation}
\label{p0Marginal}
p_0\equi{r\gg a}\frac{1}{\ln \left(\frac{r}{a}\right)}.
\end{equation}
This leads to the marginal case of Eqs.(1), (2) of the main text.

\subsection{Modification of the formalism in the case of infinite MFPT}
\label{SectionFormalismInfiniteMFPT}

We describe here how the formalism is modified in the case of processes leading to infinite mean FPTs.
This is typically the case of jump processes with broad waiting times, such as CTRWs. For these jumps process, attention must be paid in the computation of $\alpha$. As this exponent defines the potential aging of the duration of excursions (which  begin just after a jump),  $t$ and $t+\tau$ in the definition of $\alpha$ are taken \emph{just after a jump}. With this prescription, we  find that $\alpha=0$ for a CTRW (note that this contrasts with other definitions of the increments \cite{Schulz:2014}, which result in non zero values of $\alpha$).

In the compact case, all steps involved in the proof of Eqs.(1),(2) hold, and therefore   these equations are still valid in the case of infinite MFPTs.    

The case of  non-compact processes requires a separate treatment. 
 We  consider the case where the time of the $n^{th}$ excursion has a broad distribution, leading to an infinite mean FPT to the target. Specifically, we consider the case where the distribution of $\tau_n$ has a power law tail:
\begin{equation}
\label{distrib_CTRW}
P(\tau_n)\sim \frac{{\widetilde \tau}^\gamma}{\tau_n^{1+\gamma}} \frac{1}{n^{\delta}},
\end{equation}
with $\gamma\in]0,1[$ and ${\widetilde \tau}\propto R^{d_w}$. The exponent $\delta$, defined by Eq.\eqref{distrib_CTRW},  describes the aging of the durations of successive excursions. 
The key point here is to remark that $t_n=\sum_{k=1}^{n-1}\tau_k$ is a sum of now broadly distributed random variables, so that   the scalings \eqref{Expr_tn},\eqref{Expr_taun} are not valid anymore. Here the scaling of $t_n$ with $n$ is conveniently found by starting from the Laplace transform:
\begin{eqnarray}
\langle e^{-s t_n}\rangle &=&\prod_{i=1}^n\langle e^{-s \tau_i}\rangle\nonumber\\
&=& e^{\sum_{i=1}^n \ln \langle e^{-s \tau_i}\rangle}\nonumber\\
&=&e^{\sum_{i=1}^n \ln \left( 1-\frac{(s{\widetilde \tau})^\gamma}{i^\delta}+...\right)}\nonumber\\
&=&e^{-(s{\widetilde \tau}n^{\frac{1-\delta}{\gamma}})^\gamma+...},
\end{eqnarray}
which leads to
 \begin{equation}
t_n\equi{R\to\infty}R^{d_w}n^{\frac{{1-\delta}}{\gamma}}. \label{NewScalingt_n}
\end{equation}
Using  Eq.\eqref{deft}, it is  found that here also
\begin{equation}
\delta=\frac{\alpha d_w}{2}.
\end{equation}
Finally, inserting the scaling (\ref{NewScalingt_n}) into Eq.(\ref{discrete}), we find that the FPT distribution is given by Eq.(2) of the main text with $\eta=t/T_\mathrm{typ}$, but that the typical time is now
\begin{equation}
\label{generalisation}
T_\mathrm{typ}= R^{d_w}\left(\frac{R}{a}\right)^{\frac{\psi}{\gamma}\left(1-\frac{\alpha d_w}{2}\right)}. 
\end{equation}
Note that, in the case of CTRWs with infinite mean waiting times,  the formula \eqref{formulepsi} does not directly apply, because of divergences of both numerator and denominator of \eqref{rapport}.  Since $\psi$ is a purely geometrical exponent, $d_w$ in Eq.
\eqref{formulepsi} has to be replaced by the fractal dimension of the trajectory (i.e. the walk dimension of the process without waiting times).

\subsection{Explicit scalings of the moments of the FPT with the geometrical parameters}

The expressions of the distribution $G$ presented in the previous sections allow us to determine the dependence on the geometrical parameters of all the FPT moments (when they exist):
\begin{equation}
\label{moments}
\langle T^m\rangle\sim   \left \{
\begin{array}{ll}
R^{d_w(m-\theta)}r{^{d_w \theta}}& \mbox{(compact)} \\
 \frac{R^{d_w}}{\left(\ln R / a\right)^{\alpha d_w/2}} \ln\frac{r}{a}& \mbox{(marginally compact)} \\
\frac{R^{m(d_w+\psi(1-\alpha d_w/2))}}{a^{m(\psi(1-\alpha d_w/2))}} \left[1-C\left(\frac{a}{r}\right)^\psi \right]& \mbox{(non compact)} \\
\end{array}
\right ..
\end{equation}

\subsection{Derivation of the formula $\theta=1-d_f/d_w$ for the persistence exponent of processes with stationary increments (compact case)}
\label{arguetheta}

We argue here that the  relation 
\begin{equation}
\theta=1-\frac{d_f}{d_w}
\end{equation}
holds for general processes with stationary increments.

First, this relation holds for Markov processes, as can be shown from a classical renewal equation \cite{Meroz:2011}. Second, it can be recovered by comparing the result 
\begin{equation}
\langle T \rangle \sim R^{d_w(1-\theta)}
\end{equation}
obtained in the main text for compact processes (and $r$ fixed) and 
\begin{equation}
\langle T \rangle \sim R^{d_f}
\end{equation}
obtained in \cite{Guerin:2016} for non-Markovian Gaussian processes with stationary increments. Note that the results of \cite{Guerin:2016} are exact perturbatively at order $\epsilon^2$ around Markovian processes and quantitatively accurate even for strongly non-Markovian processes. Last, for FBM in 1d,  it was obtained by scaling arguments in \cite{Krug:1997} and shown  mathematically in  \cite{Molchan1999}. 

\subsection{The transience exponent $\psi$  (non compact case): derivation of Eq.(3) of the main text}

\label{six}

\subsubsection{Processes with stationary increments at long times}

\label{psistat}

\textit{Markovian processes (stationary increments).- }
For Markovian processes, the transience exponent $\psi$ can be obtained from the renewal equation. It relates the  propagator $P({\bf r},t|{\bf r'})$ (the probability density that the walker is at ${\bf r}$ at $t$ starting from ${\bf r'}$ at $t=0$) and the first-passage time density $F(t|{\bf r'})$ to reach the spherical target of radius $a$ centered at ${\bf 0}$ at time $t$:
\begin{equation}
P({\bf 0},t|{\bf r})=\int_0^t{\rm d} t' F(t'|{\bf r})P({\bf 0},t|a,t'),
\end{equation}
where $P({\bf 0},t|a,t')$ is the propagator averaged over the starting points on the sphere of radius $a$ centered at ${\bf 0}$. For a process that has stationary increments, one can write $P({\bf 0},t|a,t')$ as a function of the time lag $t-t'$ only, 
\begin{equation}
P({\bf 0},t|{\bf r})=\int_0^t{\rm d} t' F(t'|{\bf r})P({\bf 0},t-t'|a).
\end{equation}
Integrating this equation over time $t$ from $0$ to $\infty$, we obtain  the following expression of the probability $p$ to eventually reach the target: 
\begin{equation}
\label{rapport}
p=\frac{\int_0^\infty {\rm d}t P({\bf 0},t|{\bf r})}{\int_0^\infty {\rm d}t P({\bf 0},t|a)}.
\end{equation}
Using scale-invariance, the propagator can be written as \cite{D.Ben-Avraham:2000}
\begin{align}
	P({\bf 0},t|{\bf r})\sim \frac{1}{t^{d_f/d_w}}\Pi\left(\frac{r}{t^{1/d_w}}\right)
\end{align}
and we finally obtain 
\begin{equation}
p\sim\left(\frac{a}{r}\right)^{d_f-d_w},
\end{equation}
\textit{i.e.} 
\begin{equation}
\label{formulepsi}
\psi=d_f-d_w.
\end{equation}

\textit{Non-Markovian Gaussian processes with stationary increments.- }
For general non-Markovian processes, to the best of our knowledge, the exponent $\psi$ has not been studied and no exact results are available. However,  by comparing the result 
\begin{equation}
\langle T \rangle \sim R^{d_w+\psi}
\end{equation}
obtained in the main text for non-compact processes (and $r$ fixed) and 
\begin{equation}
\langle T \rangle \sim R^{d_f}
\end{equation}
obtained in \cite{Guerin:2016} for non-Markovian Gaussian processes with stationary increments, we obtain again 
\begin{equation}
\psi=d_f-d_w 
\end{equation}
for non-Markovian Gaussian processes with stationary increments.  Note that, from the very definition of $\psi$ (which involves the small target size limit), this argument can be extended to processes whose increments  are only asymptotically stationary. In addition,  we expect that the validity of this result is broader than the case of  Gaussian processes.

\subsubsection{Processes with non zero aging exponent $\alpha$.}

We now consider a process  with non zero aging exponent $\alpha$, such that:
\begin{equation}
\langle (X(t+\tau)-X(t))^2\rangle \equi{t\to\infty} t^{\alpha} \tau^{2/d_w-\alpha}.
\end{equation}
Let us define the time-changed process
\begin{equation}
X^*(t)\equiv X(t^b).
\end{equation}
Its increments satisfy on the one hand:
\begin{equation}
\langle (X^*(t+\tau)-X^*(t))^2\rangle=\langle (X((t+\tau)^b)-X(t^b))^2\rangle\sim t^{\alpha b }\left((t+\tau)^b-t^b\right)^{\frac{2}{d_w}-\alpha}\sim t^{b\alpha+(b-1)\left(\frac{2}{d_w}-\alpha\right)}\tau^{\frac{2}{d_w}-\alpha},
\end{equation}
and on the other hand by definition:
\begin{equation}
\langle (X^*(t+\tau)-X^*(t))^2\rangle \equi{t\to\infty} t^{\alpha^*} \tau^{2/d_w^*-\alpha^*}.
\end{equation}
By identification, this leads to
\begin{equation}
\alpha^*=\alpha+(b-1)\frac{2}{d_w} 
\end{equation}
and 
\begin{equation}
d_w^*=\frac{d_w}{b}.
\end{equation}
This shows that the process $X^*$ has asymptotically stationary increments if we chose $b$ with the value 
\begin{equation}
b=1-\frac{\alpha d_w}{2}.
\end{equation}
The analysis of the above paragraph then applies to $X^*(t)$ and yields
\begin{equation}
\psi_{X^*}=d_f-d_w^*=d_f-\frac{d_w}{1-\frac{\alpha d_w}{2}}.
\end{equation}
From the definition of $\psi$, it is clear that this exponent is invariant under a generic clock change of the process. As a result,
\begin{equation}
\psi_{X}=\psi_{X^*}=d_f-\frac{d_w}{1-\frac{\alpha d_w}{2}},
\end{equation}
where we have used the result of the previous subsection \ref{psistat} on processes with stationary increments at long times.



\section{Explicit results for specific stochastic processes and details of numerical simulations}

In this Section, we consider several  examples of stochastic processes, for which we determine the aging exponent $\alpha$ and the persistence exponent $\theta$, and deduce the  FPT distribution in confinement. 
We also describe the algorithms used to perform the simulations of these stochastic processes, which are used to plot Figures 2, 3 and 4 of the main text.  

\subsection{Processes with stationary increments (compact case)}

\label{compact_stat}

In the case of compact processes with stationary increments, according to Eq.(2), the distribution of the rescaled variable $\eta=T/R^{d_w}$, in the large $R$ limit, is
\begin{equation}
G(\eta;a,r,R)=h(\eta)\left(\frac{r}{R}\right)^{d_f-d_w}.\label{OLIR}
\end{equation}
where we have used the relation $\theta=1-\frac{d_f}{d_w}$ valid for processes with stationary increments, see Section \ref{arguetheta}. When the moments of the FPT exist, they are thus given by
\begin{equation}
\langle T^m\rangle\sim  R^{d_w(m-1)+d_f}r^{d_w-d_f}.  \label{ORRKE}
\end{equation}
As a particular case of this general expression, we recover the scalings with $r$ and $R$:
\begin{itemize}
\item of the case of scale-invariant Markovian processes (stationary increments). See Ref.~\cite{Condamin:2007zl} for the first moment and  Ref.~\cite{BenichouO.:2010a} for the full distribution;
\item of the mean FPT $\langle T \rangle$ in the case of a one dimensional Fractional Brownian Motion (which is a scale-invariant Gaussian non-Markovian process with stationary increments). %
\end{itemize}

\subsection{Processes with stationary increments at long times (non-compact case)}

\label{longtimes}

As stated in the main text, in the non compact case, the FPT distribution  depends only  on $d_f$, $d_w$ and $\alpha$. In turn, the dynamical exponents $d_w$ and $\alpha$ depend only on the long time asymptotics of the process.

In the case of processes with stationary increments at long times (i.e. $\alpha=0$), we have from section \ref{psistat}
\begin{equation}
 \psi=d_f-d_w.
\end{equation}
The rescaled variable is thus
\begin{equation}
\eta\equiv\frac{Ta^{d_f-d_w}}{R^{d_f}}
\end{equation}
and its distribution in the large $R$ limit is given by 
\begin{equation}
\label{NCstat1}
G(\eta;a,r,R)=h(\eta)\left[1-C\left(\frac{a}{r}\right)^{d_f-d_w}\right].
\end{equation}
When the moments of the FPT exist, they are thus given by
\begin{equation}
\label{NCstat2}
\langle T^m\rangle \sim \left(\frac{R^{d_f}}{a^{d_f-d_w}}\right)^m \left[1-C\left(\frac{a}{r}\right)^{d_f-d_w}\right].
\end{equation}
As particular cases of this general expression, we recover the scalings with $r$ and $R$  of  scale-invariant Markovian processes  (see \cite{Condamin:2007zl} for the first moment and  \cite{BenichouO.:2010a} for the full distribution).
 
Note that these scalings hold in the important case of a $d$-dimensional  Fractional Brownian Motion (for both equilibrated and non-equilibrated initial conditions), such that $H<1/d$, with $d_f=d$ (in order to have a non compact process), see also below.

\subsection{Fractional Brownian Motion with equilibrated or non-equilibrated  initial conditions}

\label{QuenchedFBMSection}

\subsubsection{Theoretical results}
The 1d Fractional Brownian Motion (FBM) is the Gaussian process with constant mean (here we set this mean to $X_0$) and correlations $\langle (X(t)-X_0)(X(t')-X_0)\rangle=K[ t^{2H}+t'^{2H}- | t-t'|^{2H}]$. $H$ is the Hurst exponent ($0<H<1$), and here we take $K=1$. We define the $d$-dimensional FBM as $\ve[X](t)=(X_1(t),...,X_d(t))$ where the $X_i(t)$ are independent one dimensional FBMs. The FBM is a non-Markovian process with stationary increments, the results of the former sections therefore apply with a walk dimension $d_w=1/H$:  the FPT statistics in a large confining volume follow Eqs. (\ref{OLIR}),(\ref{ORRKE}) in the compact case ($d<1/H$) and Eqs. (\ref{NCstat1}),(\ref{NCstat2}) in the non-compact case ($d>1/H$). We tested both cases in our simulations (see below for details).

Since $X(t)$ is a non-Markovian process, its statistics  depend on the one of   trajectories in the past ($t<0$). To specify the ``initial state'', it is useful to consider a microscopic model which is equivalent to a FBM in  specific limits. We introduce the stochastic process $X(t)$ defined as  the local height of a fluctuating interface $h(x,t)$ at a given position, $X(t)=h(x_0,t)$. 
Following Ref. \cite{Krug:1997}, we  assume the following dynamics
\begin{equation}
\frac{\partial h}{\partial t}=-(-\nabla ^{2})^{\frac{1}{2-4H}} \, h + \xi(x,t), \label{EqInterface}
\end{equation}
where $\xi(x,t)$ is a Gaussian white noise, satisfying $\langle\xi(x,t)\xi(x',t')\rangle=\delta(t-t')\delta(x-x')$. 
$X(t)$ is then a Gaussian process. It has been shown that the persistence exponent of $X(t)$ is then strongly dependent on the initial interface height distribution \cite{Krug:1997}.

If the interface is initially equilibrated, then $\langle X(t)X(t')\rangle=t^{2H}+t'^{2H}- | t-t'|^{2H}$ and $X(t)$ is a Fractional Brownian Motion with Hurst exponent $H$ and has stationary increments. If the interface is initially flat, $h(x,t=0)=0$ for all $x$, then the process $X(t)$ becomes non-Markovian with non-stationary increments, and the correlator reads
\begin{equation}
\langle X(t)X(t')\rangle=(t+t')^{2H}- | t-t'|^{2H}\label{QuenchedFBMCov}
\end{equation}
In this case, we call the process $X(t)$  an ``initially quenched'' Fractional Brownian Motion. Its exponent $\theta$ is not known analytically, but can be determined numerically. In Fig. 2A of the main text, Eq. (4) of the main text for the mean FPT is checked in the particular case $H=3/8$, in which case  one can approximate $\theta(H=3/8)=0.84...$ \cite{Krug:1997}. In this case, $\left \langle T \right \rangle $ is sublinear in $V$.
In order to obtain an example of process with  $\left \langle T \right \rangle $ superlinear in $V$, we consider the process defined by Eq.~(\ref{QuenchedFBMCov}) with $H=0.65$. In this case, $\theta\simeq 0.3$ (see Ref.~\cite{Krug:1997}). Note that this process cannot be easily matched on the evolution of a point of an interface, contrarily to the case $H=3/8$.

\subsubsection{Description of simulations}
\begin{figure}[h]
\includegraphics[scale=0.43]{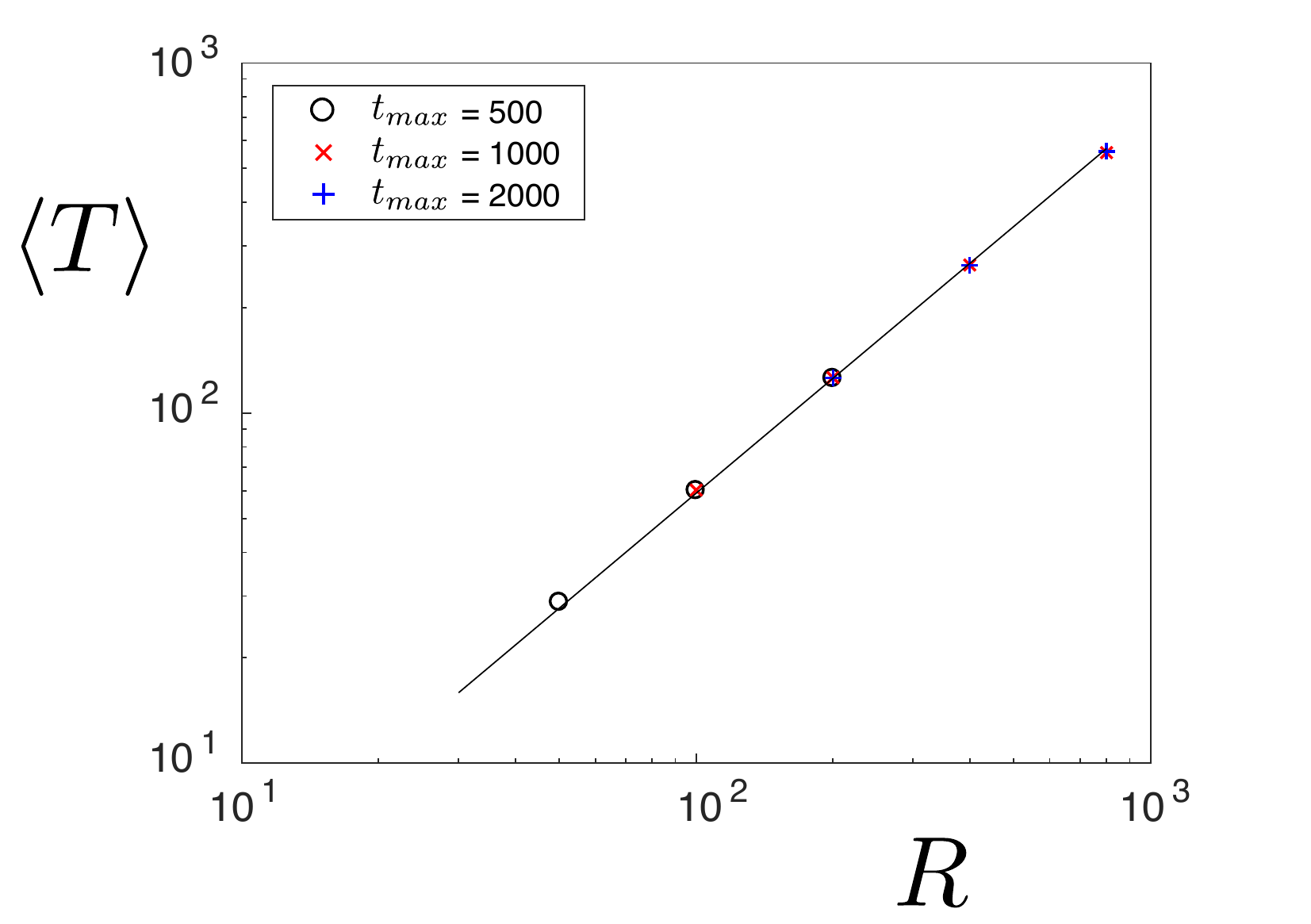} 
\caption{\textbf{Control of the validity of the estimator (\ref{EstimatorT}) for the MFPT in simulations of the 1d initially quenched FBM, with $H=0.65$.(see text)} }
\label{checkT}
\end{figure}

\begin{enumerate}
\item \textbf{One dimensional Fractional Brownian Motion (FBM)}. The algorithm used to sample the 1d FBM trajectories was the circulant matrix algorithm (also called the Davies and Harte method) \cite{wood1994simulation,davies1987tests,dietrich1997fast}. 
This method generates trajectories $X(t_i)$ with a constant time step $\Delta t=t_{i+1}-t_i$, until a fixed maximal time $t_\mathrm{max}$, with a number of operations of the order of $N\ln N$, with $N= t_\mathrm{max}/\Delta t$. The value of $t_\mathrm{max}$ was chosen so that the number of trajectories that did not reach the target was negligible. The time step $\Delta t$ was then reduced until obtaining convergence, with a FPT density that do not depend on $\Delta t$. In Fig. 2A and Fig. 3A, for $H=0.4$, we used $t_\mathrm{max}=524888$ and $\Delta t=0.1$, except for $R=320$ where $t_\mathrm{max}$ was doubled, with the same time step. The initial position is set to $X_0=1$. It was checked in Ref. \cite{Guerin:2016} that defining reflecting boundaries with the Hosking algorithm does not change the mean FPT in the large volume limit. In Fig. 2A, the mean FPT and the volume $V=R$ are rescaled by the factors $(V_0; T_0) = (20; 27.2)$.

\item \textbf{Two dimensional Fractional Brownian Motion.}
The $2$-dimensional FBM  is defined as  $\ve[X]=(X_1(t),X_2(t))$ where the $X_1(t),X_2(t)$ are
$1$-dimensional independent  FBMs. $X_1(t),X_2(t)$ were simulated with the circulant matrix algorithm described above. The target radius was set to $a=1$, and the domain defined as the square of size $\sqrt{\pi} R$ with periodic boundary conditions. In Fig. 3B, for $H=0.35$, we used $t_\mathrm{max}=2\times10^6$ and a time step $\Delta t=0.1$. In Fig. 3D and Fig. 2B, for $H=0.7$, we used  $t_\mathrm{max}=10^6$ and $\Delta t=0.1$, and the initial distance to the target was $r=10$. In Fig. 2B, the results were rescaled by the factors $(V_0; T_0) = (50; 155.4)$.

\item \textbf{One dimensional FBM with non-equilibrated initial conditions and $H=3/8$}. We consider here the process defined as $X(t)=h(0,t)$, where $h(x,t)$ is the height of an interface. As discussed in Section \ref{QuenchedFBMSection}, $X(t)$ is a FBM only when the initial distribution of $h(x,t)$ is the equilibrium distribution.  Here we consider that the interface is initially flat, $h(x,t=0)=X_0$. In Fig. 2A, $X_0=10$. The simulations are performed by the stochastic integration of the Langevin equation (\ref{EqInterface}) following the algorithm of Ref. \cite{Krug:1997}, with  $H=3/8$. The interface is described by its height at a finite number $N$ of positions, $h_i(t)=h(i \Delta x,t)$. Here, we take $\Delta x=1$ and $N=200$ discrete positions, and we use periodic boundary conditions ($h_0=h_N$). 
The time-step was set to $\Delta t=0.1$, and the convergence of the results with the time-step has been checked. The maximal time was  $t_\mathrm{max}=4.10^{7}$. It was checked that the relaxation time of the system  (which grows as $N^4$ for $H=3/8$) was much larger than   $t_\mathrm{max}$.  In Fig. 2A, the results were rescaled by the factors $(V_0; T_0) = (160; 91.3)$.

\item \textbf{One dimensional initially quenched FBM, $H=0.65$}. 
We consider here the simulations of a process $X(t)$ starting at $X_0=r=1$ with covariance  given by Eq.(\ref{QuenchedFBMCov}). The targets are at the positions 0 and $R$. We simulated the trajectories at $N$ discrete times $t_n=n\Delta t$ by using a Cholesky decomposition of  the covariance matrix $\Sigma_{ij}\equiv \langle X(t_i)X(t_j)\rangle$, which can be written as $\Sigma=LL^{t}$, with $L$ a lower triangular matrix. 
Then, a trajectory is generated by computing $X(t_i)=\sum_{j=1}^NL_{ij}u_j$, where the $u_j$ are independent Gaussian variables of zero mean and variance unity. The time-step was set to $\Delta t =0.03$ to ensure convergence. 
Because of the large memory requirement of this algorithm (it increases as $N^2$), we could not chose $t_\mathrm{max}=N\Delta t$ large enough to obtain a negligible proportion $q$ of trajectories which did not find the target. In order to avoid errors in the measurement of the MFPT, we proceeded as follows. Observing that the tail of the FPT distribution is exponential, we fitted the tail of the survival probability with $S(t)\sim B (R/r)^{\theta/H} e^{-a t/R^{1/H}}$. Then, calling $T_{\mathrm{emp}}$ the MFPT restricted to successful runs (that is with a FPT smaller than $t_\mathrm{max}$), the MFPT was estimated with the formula
\begin{equation}
\left \langle T \right \rangle = (1-q) T_\mathrm{emp} +aB R^{(\theta-1/H}e^{-a t_\mathrm{max}/R^{1/H}}(1+a t_\mathrm{max}/R^{1/H}) \label{EstimatorT}
\end{equation}
We checked the validity of this procedure by ensuring that different $t_\mathrm{max}$ lead to the same result for the MFPT (see Fig.\ref{checkT}). In Fig. 2A, the results were rescaled by the factors $(V_0; T_0) = (50; 28.8)$. In Fig. 3C, we used $t_\mathrm{max}=500$ for $R=50$, $t_\mathrm{max}=1000$ for $R=100, 200$ and $t_\mathrm{max}=2000$ for $R=400$.

\end{enumerate}

\subsection{The random acceleration process and its generalizations}

 \subsubsection{Theoretical results}
We consider here the $d$-dimensional process $\ve[X](t)=(X_1(t),...,X_d(t))$ of order $n$ ($n\ge 2$), defined by
\begin{equation}
\frac{{\rm d}^n}{{\rm d}t^n}X_i(t)=  \xi_i(t), \label{dDimRAP}
\end{equation}
where $\xi_i(t)$ is a Gaussian white noise with zero mean satisfying $\langle\xi_i(t)\xi_j(t')\rangle=K\delta(t-t')\delta_{ij}$, with $K$ a coefficient that can be set to unity by appropriate rescaling, $K=1$. $\ve[X](t)$ is then a non-Markovian process, with non stationary increments. For $n=2$, $X_i(t)$ is a random acceleration process; for $n>2$ one obtains higher order processes. We will in particular consider the case $n=3$ (random jerk process). 

We first investigate the aging features of this class of processes. Denoting by $x$ a generic coordinate $X_i$  and  using the representation $x(t)=\int_{0}^{t} | t-t' |^{n-1} \xi (t') dt'$, we compute easily the autocorrelation function :
\begin{align}
\left \langle x(t+\tau)x(t) \right \rangle =& \int_{0}^{t+\tau} dt_{1} \int_{0}^{t} dt_{2} | t+\tau-t_{1} |^{n-1} |t-t_{2}|^{n-1} \delta (t_{1}-t_{2})  =t^{2n-1}\int_{0}^{1} x^{n-1} (x+(\tau/t))^{n-1} dx.
\end{align}
Hence the increments are given by :
\begin{equation}
\left \langle (x(t+\tau)-x(t))^{2} \right \rangle=(t+\tau)^{2n-1}\int_{0}^{1}  x^{2n-2} \, dx \;-2t^{2n-1} \int_{0}^{1} x^{n-1} (x+(\tau/t))^{n-1}\, dx \;+t^{2n-1} \int_{0}^{1}  x^{2n-2} \, dx
\end{equation}
It is straightforward to see that both terms proportional to $t^{2n-1}$ and $t^{2n-2}$ do cancel. Finally we get 
\begin{align}
\left \langle (x(t+\tau)-x(t))^{2} \right \rangle \propto t^{2n-3} \tau^{2}. 
\end{align}
The walk dimension $d_w$ and the aging exponent $\alpha$ are then given by (for $n\ge2$)
\begin{equation}
d_w(n)=\frac{2}{2n-1},\hspace{2cm}
\alpha(n)=2n-3,
\end{equation}
where we have used the definition (see main text)
\begin{equation}
\langle (X(t +\tau )-X(t ))^2\rangle \sim_{t \to\infty}  t^{\alpha} \tau^{2/d_w-\alpha}.
\end{equation}
As a consequence, the transience exponent is 
\begin{equation}
\psi=d-1.
\end{equation}
For $d>1$, $\psi>0$ and the process is therefore non-compact, whereas it is compact if $d=1$. 

\begin{figure}[h]
\includegraphics[scale=0.38]{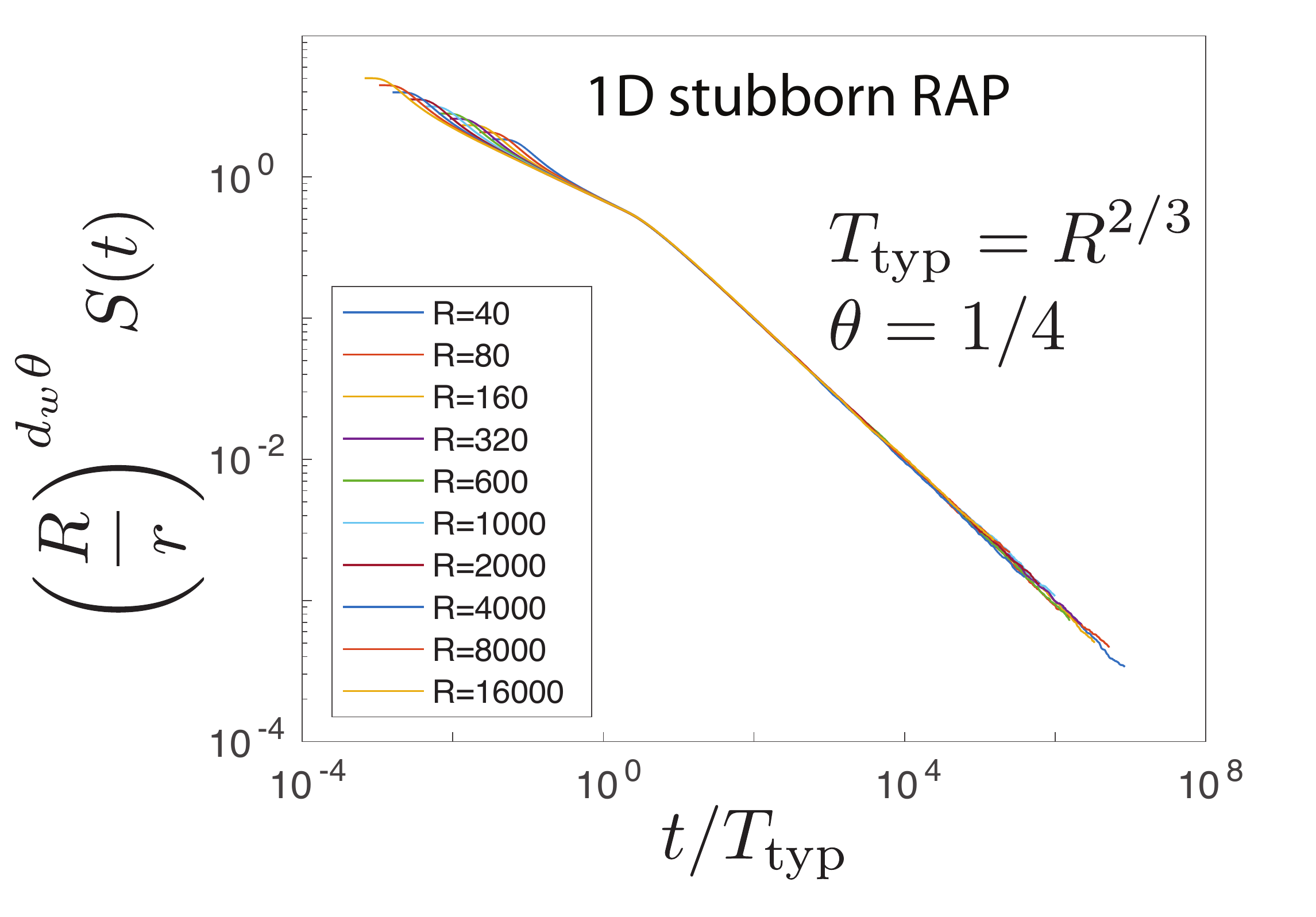} 
\caption{\textbf{Validity of the scalings of the FPT distribution with geometrical parameters for the stubborn RAP.} }
\label{Figstubborn}
\end{figure}

Our theoretical results then predict that
\begin{align}
	T_\mathrm{typ}=
\begin{cases}
R^{\frac{2}{2n-1}} & (d=1)\\
R^{\frac{2}{2n-1}}\left(\frac{R}{a}\right)^{\frac{2(d-1)}{2n-1}} & (d>1), 
\end{cases}
\end{align}
and $\eta=T/T_\mathrm{typ}$ is asymptotically distributed in the limit of large volume according to
 \begin{align}
	G(\eta;a,r,R)=
\begin{cases}
h(\eta)\left(\frac{r}{R}\right)^{\frac{2\theta}{2n-1}}& (d=1)\\
h(\eta)\left[1-C\left(\frac{a}{r}\right)^{d-1}\right]. & (d>1).
\end{cases}\label{G_RAP}
\end{align}
In the compact case $d=1$, the persistence exponent is known either exactly (for $n=2$) or approximately (for $n\ge3$). It is also known in the case of a partially absorbing target, where the random walker has a finite probability $q$ to actually find the target at each visit:
\begin{align}
\theta=\begin{cases}
1/4 & (n=2)\\
0.2202... & (n=3)\\
\frac{1}{4}\left[1-\frac{6}{\pi}\sin^{-1}\left(\frac{1-q}{2}\right)\right]  & (n=2, \text{partial absorption with probability  } q).
\end{cases}
\end{align}
Hence, the mean FPT, obeys the scaling (when it exists)
\begin{align}
	\langle T\rangle\sim\begin{cases}
R^{\frac{2(1-\theta)}{2n-1}}r^{\frac{2(\theta)}{2n-1}} & (d=1)\\
\frac{R^{\frac{2d}{2n-1}}}{a^{\frac{2(d-1)}{2n-1}}}\left[1-C\left(\frac{a}{r}\right)^{d-1}\right]& (d>1)
\end{cases}
\end{align}
and the  higher moments can be obtained by using Eq.~(\ref{moments}). 
Note  that, in particular, in the case $n=2, d=1$, one has $\langle T \rangle \sim r^{1/6} R^{1/2}$, which is in agreement with  Ref.~\cite{Masoliver:1996} in the case of reflecting boundary conditions (where the sign of the velocity is immediately reversed when the confining boundary is reached $v\to -v$). We also considered another type of boundary conditions, with a confining boundary that stops the particle while its ``target'' velocity keeps evolving according to Eq.~(\ref{dDimRAP}) until it changes  sign and the particle escapes from the boundary. We call the RAP with this kind of boundary conditions the ``stubborn'' random acceleration process. In this case the moments of the FPT distributions are infinite, but the distribution of FPT still satisfies the scaling \eqref{G_RAP} (Fig. \ref{Figstubborn})

 \subsubsection{Simulation details}
We now describe  the simulations of the random acceleration process and its generalizations that were used in Figures 2,3 and 4 of the main text
\begin{itemize}
\item \textbf{Simulations of the Random Acceleration Process (n=2)}. The stochastic trajectories are generated by integrating numerically the Langevin equation (\ref{dDimRAP}) with the algorithm introduced in \cite{Bicout:2000}. This algorithm does not generate any error due to the time discretization, as it creates the exact pdf for the joint density position-velocity. We need however to choose a time step $\Delta t$ that is small enough to ensure that the target is not missed during one step. Since the velocity typically grows as $\sqrt{t}$, we have to reduce the time-step at each iteration, in order to keep $\Delta x \sim v \Delta t$ small. A simple way to do that is to take the time step of the $n^\mathrm{th}$ step to be $(\Delta t)_{n}=D/n^{1/3}$, with a small enough $D$. With this choice, the spreading of the velocity does not increase with $n$. The value of $D$ is decreased up to reach convergence of the results. We chose $D=0.7$ in all cases. The initial velocity is $v_0=0$. Additional details are:
\begin{enumerate}
\item In the case of reflecting conditions in $1d$,  the reflecting boundary, located at $x=R/2$, can be replaced by a second target at $x=R$. The initial position is set to $X_0=1$. The probability $q$ of being absorbed at each target crossing event is set to  $q=1$ or $q=0.7$. In Fig. 2A, the mean FPT and the volume $V=R$ are rescaled by the factors $(V_0; T_0) = (40; 8.05)$ in the case $q=1$ and $(V_0; T_0) = (20; 8.79)$ when $q=0.7$.  
\item In $1d$, we also considered the  ``stubborn'' acceleration process; when reaching the confining boundary, the velocity at step $i$ evolves according to $v_{i+1}=v_i+\sqrt{(\Delta t)_i}$ while the position remains constant at $x_i=R/2$ until reaching a negative velocity that drives the walker away from the boundary. The results for this process  are presented in Fig. \ref{Figstubborn}. 
\item In $2$ ($3$) dimensions, we consider a square (cubic) confining volume $V=R^d$ which, for reflecting boundary conditions, can be replaced by an infinite periodic array of targets. The initial distance to the target is taken to be $r=10$, the target radius is $a=1$. In Fig. 2B, the mean FPT and the volume $V$ are rescaled by the factors $(V_0; T_0) = (100; 556.4)$ (in 2d) and and $(V_0; T_0) = (100; 14870)$ (in 3d).
\end{enumerate}
\item \textbf{Simulations of the Random Jerk Process (n=3)}. The random Jerk process is defined by Eq.~(\ref{dDimRAP}), with $n=3$. The trajectories are sampled by using $X(t_{n+1})=X(t_n)+Y(t_n)( t_{n+1}-t_n)$, where $Y_n$ is a Random Acceleration Process (see above). In order to keep the velocity small, we chose $(\Delta t)_{n}=D/n^{3/5}$, and we reach convergence for $D=0.1$. In $1d$, the initial distance to the target is $X_0=1$, and the domain size $R/2$. In $2d$, the initial distance to the target is $r=10$, the target radius is $a=1$, and the domain is  a square of size $LR\times R$ with period boundary conditions. In Fig.2, the mean FPT and the volume $V$ are rescaled by the factors $(V_0; T_0) = (160; 7.6)$ (in 1d, Fig. 2A) and = $(V_0; T_0) = (40; 24.7)$ (in 2d, Fig. 2B).
\end{itemize}

\subsection{The case of L\'evy flights}

We now consider the case of L\'evy flights, where at each step a $d$-dimensional  walker  performs in a random direction a jump whose length $l$ is drawn from a distribution with power-law tail:
\begin{equation}
p(l)\sim \frac{1}{l^{1+\beta}}.
\end{equation}
These processes have been repeatedly invoked in the literature on random search processes \cite{viswanathan2011physics}. Two definitions of the first-passage time to the target have been used in previous works. In the first definition, the target can be detected only when the walker changes direction. This first-passage time will be termed here  a \textit{first-arrival time}  \cite{al:2003,1751-8121-44-25-255003}. In the second definition, the target can be detected as soon as  crossed by the trajectory of the  walker. This first-passage time will be termed here  a \textit{first-crossing time}.

We consider only the case $0<\beta<2$ (note that for $\beta>2$, the process converges  to Brownian motion). The dimension of the walk is  given by:
\begin{equation}
d_w=\beta. 
\end{equation}

\subsubsection{First arrival case}

In this case, the renewal type argument of section  \ref{arguetheta} holds. For integer dimensions, the process is compact for $d=1$ and non-compact for $d\ge 2$. In the compact case, we have 
\begin{equation}
\theta=1-\frac{d_f}{d_w}=1-\frac{1}{\beta}.
\end{equation}
and the aging exponent is $0$. 
Thus, we predict that the typical time is, for any $\beta\in ]0,2[$
\begin{align}
T_\mathrm{typ}=
\begin{cases}
R^\beta & (d=1)\\
\frac{R^{d}}{a^{d-\beta}} &(d=2,3)
\end{cases}
\end{align}
and that the distribution of the rescaled FPT $\eta=T/T_\mathrm{typ}$ is asymptotically distributed according to
\begin{equation}
G(\eta;a,r,R)=
\begin{cases}
h(\eta)(r/R)^{\beta-1} & (d=1)\\ 
h(\eta)\left[1-C\left(\frac{a}{r}\right)^{d_f-\beta }\right]. &(d=2,3...)
\end{cases}\label{DistriGFirstArrival}
\end{equation}
The results of Section \ref{compact_stat} apply with $d_f=1$ and $d_w=\beta$. They are consistent with the Markovian prediction of reference \cite{BenichouO.:2010a}.

\subsubsection{First crossing}

In this case, the renewal type argument of section  \ref{arguetheta} does not apply directly. In one dimension, the process is compact, and the persistence exponent can  be obtained from the Spare Anderson theorem \cite{Redner:2001,Bray:2013}
\begin{equation}
\theta=1/2.
\end{equation}
The rescaled variable $\eta=T/R^{\beta}$ is  thus asymptotically distributed in the large $R$ limit according to the distribution 
\begin{equation}
G(\eta;a,r,R)=h(\eta)\left(\frac{r}{R}\right)^{\beta/2}.
\end{equation}
The moments are then given by
\begin{equation}
\langle T^m\rangle=  R^{\beta(m-1/2)}r^{\beta/2}.
\end{equation}
Note that the scalings with $r$ and $R$ of   the first-moment obtained in  \cite{Buldyrev:2001} are recovered from this general expression of the moments.

In the non-compact case, when $\beta\in]1,2[$, $d_w=\beta$ and $\psi=d_f-\beta$. Thus in this case, the scaling results are the same for both arrival and crossing prescriptions, Eq.~(\ref{DistriGFirstArrival}). 

The situation is different for $\beta\in]0,1[$. In this case, the trajectories are the same as for a L\'evy Walk, for wich it is argued  below that  $\psi=d_f-1$. Note that Eq. (3) of the main text applies for first-passage problems, and not directly to first-crossing problems. 
Finally,
\begin{equation}
T_\mathrm{typ}\sim \frac{R^{d_f+\beta-1}}{a^{d_f-1}}
\end{equation}
and
\begin{equation}
G(\eta;a,r,R)=h(\eta)\left[1-C\left(\frac{a}{r}\right)^{d_f-1 }\right].
\end{equation}
Notably, in this case, the scaling results are therefore different  for  arrival and crossing prescriptions.

\subsection{The case of L\'evy walks}

\label{LWcompact}

L\'evy walks provide a  natural physical generalization of L\'evy flights in which the instantaneous velocity of the walker is bounded, as opposed to L\'evy flights. We set here this instantaneous velocity to unity. A $d$-dimensional  L\'evy walker performs a series of independent  and randomly oriented  ballistic excursions at constant speed, whose length $l$ is  drawn from a distribution with power-law tail:
\begin{equation}
p(l)\sim \frac{1}{l^{1+\beta}}.
\end{equation}
For $\beta\in]0,1[$, the process is known to be scale-invariant \cite{Zaburdaev:2015}, so that our results for the FPT apply directly. For $\beta\in]1,2[$, even if the process is not scale-invariant \cite{Zaburdaev:2015}, it is known that the bulk of the propagator is scale-invariant. Knowing that the weight of the ballistic fronts that compose the tail of the distribution is negligible, these tails are irrelevant for the determination of the FPT statistics, and we therefore make use of the scalings of the bulk.

Let us first consider the compact situation, obtained for a one-dimensional case.
For $\beta\in]0,1[$, $d_w=1$ and $\theta=\beta/2$ (see  Ref.~\cite{Korabel:2011,Bray:2013}) while for $\beta\in]1,2[$, $d_w=\beta$ (see Ref.~\cite{Zaburdaev:2015}) and $\theta=1/2$. 
The typical time is therefore
\begin{align}
T_\mathrm{typ}=
\begin{cases}
R &(0<\beta<1, d=1)\\
R^\beta & (1\le\beta<2, d=1) 
\end{cases}
\end{align}
and, for  $\beta\in]0,2[$, the rescaled variable $\eta=T/R$ is asymptotically distributed in the large $R$ limit according to the distribution 
\begin{equation}
G(\eta;a,r,R)=h(\eta)\left(\frac{r}{R}\right)^{\beta/2}.
\end{equation}

In the non-compact case, for  $\beta \in]1,2[$, $d_w=\beta$ (see Ref.~\cite{Zaburdaev:2015}) and  $\psi=d_f-\beta$. 
In this case, 
\begin{equation}
T_\mathrm{typ}\sim \frac{R^{d_f}}{a^{d_f-\beta}}
\end{equation}
and
\begin{equation}
G(\eta;a,r,R)=h(\eta)\left[1-C\left(\frac{a}{r}\right)^{d_f-\beta}\right].
\end{equation}

Next, for $\beta \in]0,1[$, $d_w=1$ (see Ref.~\cite{Zaburdaev:2015}) and  $\psi=d_f-1$. In this case, the scaling of $T_\mathrm{typ}$ actually depends on the nature of the possible directions of the velocity of L\'evy walks. We consider here two natural choices, as defined in Ref.~\cite{2016arXiv160502908Z}: (i) in the XY model, the particle is allowed to move only on one axis at a time (implying in particular that the walker in confinement can be trapped in long lasting periodic trajectories) ; (ii) in the uniform model, at each reorientation point the particle chooses a random direction of motion specified by an angle uniformly distributed in $[0,2\pi]$ (in this case, a long enough trajectory typically finds the target). 

(i) In the XY model, durations  of excursions (as defined in the main text) are broadly distributed. From \eqref{generalisation}, it is found that 
\begin{equation}
T_\mathrm{typ}\sim \frac{R^{1+(d_f-1)/\beta}}{a^{(d_f-1)/\beta}}
\end{equation}
and
\begin{equation}
G(\eta;a,r,R)=h(\eta)\left[1-C\left(\frac{a}{r}\right)^{d_f-1}\right].
\end{equation}

(ii) In the uniform model, durations  of excursions (as defined in the main text) have a finite first moment. Thus,
\begin{equation}
T_\mathrm{typ}\sim \frac{R^{d_f}}{a^{d_f-1}}
\end{equation}
and
\begin{equation}
G(\eta;a,r,R)=h(\eta)\left[1-C\left(\frac{a}{r}\right)^{d_f-1}\right].
\end{equation}


Concerning the simulation details, at each step we generate the size $l$ of the next jump from the distribution $p(l)$. It is  defined by its Fourier transform $\tilde{p}(k)=\exp(-k^{\beta})$. The time is incremented by one at each jump.

In $2d$,  in the XY convention, we first choose randomly a direction ($x$ or $y$ with a probabilty 1/2) and then perform a jump. In the angular convention, we first randomly choose a direction, that is an angle in $[0,\pi]$ and then perform a jump along this direction.
In the $1d$ simulation, we take $r=1$ and the domain size   $V=R$. In the $2d$ case, the radius of the target is  set to $a=1$, the initial position $r=10$, and targets are  distributed on a periodic lattice of mesh-size $R$, so that $V=R^{d_{f}}$. In Fig. 2B of the main text, we used $(V_0; T_0)= (100; 14180)$ in the case of XY two-dimensional L\'evy Walks and $(V_0; T_0) = (200; 19540)$ in the case of XY two dimensional L\'evy Flights.

\subsection{Scaled processes}

In this section, we start with a scale-invariant stochastic process $X^{(0)}(t)$ on a finite domain of fractal dimension $d_f$, with vanishing aging exponent $\alpha=0$ and with stationary increments, so that
\begin{equation}
\langle (X^{(0)}(t+\tau)-X^{(0)}(t))^2\rangle \propto \tau^{2/d_w^{(0)}},
\end{equation}
and we introduce the so-called \textit{scaled process} $X(t)$, defined by
\begin{equation}
X(t)\equiv X^{(0)}(t^\beta),
\end{equation}
with $\beta>0$. For example, the Scaled Brownian motion, which corresponds to the particular case where $X(t)$ is the Brownian motion (and thus with $d_w^{(0)}=2$),  has been   used to model anomalous diffusion of passive tracers in complex and biological systems \cite{Safdari:2015}. In what follows, we determine explicitly the typical time $T_{\rm typ}$ and the distribution $G$ of this scaled process in terms of $d_f$ and $d_w^{(0)}$ in two ways : (i) by applying directly our results on the FPT distribution to the scaled process $X(t)$ ; (ii) by deducing them from our results on the FPT distribution of the starting process $X^{(0)}$.

Let us first apply  our results on the FPT distribution to the scaled process $X(t)$.  
The increments of the scaled process can be written as
\begin{eqnarray}
\langle (X(t+\tau)-X(t))^2\rangle &=&\langle (X^{(0)}((t+\tau)^\beta)-X^{(0)}(t^\beta))^2\rangle\nonumber\\
&\equi{t\to\infty}&((t+\tau)^\beta-t^\beta)^{2/d_w^{(0)}}\nonumber\\
&\equi{t\to\infty}&t^{2(\beta-1)/d_w^{(0)}}\tau^{2/d_w^{(0)}}.
\end{eqnarray}
Thus, for the scaled process, the aging exponent $\alpha$ and the walk dimension $d_w$ are given by
\begin{equation}
\alpha=\frac{2(\beta-1)}{d_w^{(0)}}, \hspace{2cm}
d_w=\frac{d_w^{(0)}}{\beta},\label{alphaScaled}
\end{equation}
where we have used the definition
\begin{equation}
\langle (X(t+\tau)-X(t))^2\rangle \propto t^{\alpha}\tau^{2/d_w-\alpha}
\end{equation}
given in the main text.
The transience exponent takes the value 
\begin{equation}
\psi=d_f-d_w^{(0)}=\psi^{(0)},\label{psiScaled}
\end{equation}
and we thus see that the scaled process is non-compact if and only if the original process is also non-compact. 

In addition, knowing that the survival probability in infinite space (defined as the probability that the target has not been reached at time $t$) can be written as $S_\infty(t)=S_\infty^{(0)}(t^\beta)$, we have
\begin{equation}
\theta=\beta\theta^{(0)}=1-\frac{d_f}{d_w^{(0)}}, \label{ThetaScaled}
\end{equation}
where the last equality follows from Section \ref{arguetheta}. 
Finally,  for the scaled process 
\begin{equation}
T_{\rm typ}=
\begin{cases}
R^{d_w^{(0)}/\beta}, & (\text{if } d_w^{(0)}>d_f, \text{ compact case}) \\
\left(\frac{R^{d_f}}{a^{d_f-d_w^{(0)}}}\right)^{1/\beta} & (\text{if } d_w^{(0)}<d_f, \text{ non-compact case})
\end{cases}
\end{equation}
and the distribution of the rescaled FPT variable, $\eta=T/T_\mathrm{typ}$, reads [see Eq.(2)]
\begin{equation}
G(\eta;a,r,R)=
\begin{cases}
h(\eta)\left(\frac{r}{R}\right)^{d_w^{(0)}-d_f} & (\text{if } d_w^{(0)}>d_f),\\
h(\eta)\left[1-C\left(\frac{a}{r}\right)^{d_f-d_w^{(0)}}\right]&(\text{if } d_w^{(0)}<d_f)
\end{cases}
\label{method1}
\end{equation}
where we have used the values of the exponents given by Eqs. \eqref{alphaScaled}, \eqref{psiScaled}, \eqref{ThetaScaled}.

It is instructive to recover this result by starting from the  results on the FPT distribution of the original (non-scaled)  process $X^{(0)}$. 
According to Eq.(1), if we define 
\begin{equation}
T_{\rm typ}^{(0)}=
\begin{cases}
R^{d_w^{(0)}}, & (\text{if } d_w^{(0)}>d_f) \\
\left(\frac{R^{d_f}}{a^{d_f-d_w^{(0)}}}\right) & (\text{if } d_w^{(0)}<d_f)
\end{cases}
\end{equation}
 the distribution of the rescaled FPT  $\eta^{(0)}=T/T_{\rm typ}^{(0)}$ is asymptotically given by 
\begin{equation}
G^{(0)}(\eta^{(0)};a,r,R)=
\begin{cases}
h^{(0)}(\eta^{(0)})\left(\frac{r}{R}\right)^{d_w^{(0)}-d_f} & (\text{if } d_w^{(0)}>d_f),\\
h^{(0)}(\eta^{(0)})\left[1-C\left(\frac{a}{r}\right)^{d_f-d_w^{(0)}}\right]&(\text{if } d_w^{(0)}<d_f)
\end{cases}
\end{equation}
The FPT distribution $F(t)$ of the scaled process can be deduced from $F^{(0)}$ as follows:
\begin{eqnarray}
F(t)&=&-\frac{\rm d}{{\rm d}t}S(t)\nonumber\\
&=&-\frac{\rm d}{{\rm d}t}S^{(0)}(t^\beta)\nonumber\\
&=&\beta t^{\beta-1}F^{(0)}(t^\beta).
 \end{eqnarray}
Therefore, the FPT distribution $G(\eta)$ of the rescaled variable $\eta\equiv t/R^{d_w}$ corresponding to the scaled process is given by
\begin{eqnarray}
G(\eta)&=&T_\mathrm{typ}F(t)\nonumber\\
&=&T_\mathrm{typ}\beta t^{\beta-1}F^{(0)}(t^\beta)\nonumber\\
&\sim&\frac{T_\mathrm{typ}}{T_\mathrm{typ}^{(0)}} t^{\beta-1} h^{(0)}\left(\frac{t^{\beta}}{(T_\mathrm{typ}^{(0)})^\beta}\right)\times\begin{cases}
\left(\frac{r}{R}\right)^{d_w^{(0)}-d_f} & (\text{if } d_w^{(0)}>d_f),\\
\left[1-C\left(\frac{a}{r}\right)^{d_f-d_w^{(0)}}\right]&(\text{if } d_w^{(0)}<d_f)
\end{cases}
\end{eqnarray}
Noting that $T_\mathrm{typ}^\beta=T_\mathrm{typ}^{(0)}$, the result \eqref{method1} is recovered.

\subsection{The case of Continuous time random walks (CTRWs)}

\label{CTRWcompact}

In this section, we consider a CTRW on a finite domain of fractal dimension $d_f$: the random walker moves on a network of fractal dimension $d_f$, at each time step a neighboring site is chosen at random, and the  waiting time $t$ at a given site is drawn from a given distribution $\rho(t)$. For the sake of simplicity, we limit ourselves to the important case of a broad distribution of waiting times:
\begin{equation}
\rho(t)\equi{t\to\infty}\frac{1}{t^{\beta+1}},
\end{equation}
with $\beta\in]0,1[$. Note that in this case, the mean waiting time at each site is infinite, which also implies that the moments of the FPT to a target are infinite. 
With the prescription given in  \ref{SectionFormalismInfiniteMFPT}, the aging exponent of such a walk is $\alpha=0$. 

We introduce the corresponding discrete time process $X_n$, representing the walker position after the $n^\mathrm{th}$ step. $X_n$ can be seen as a random walk where  the walker jumps at all units of time (instead of jumping after a random waiting time). We  denote by $d_w^{(0)}$ its walk dimension:
\begin{equation}
\langle X_n^2\rangle \sim n^{2/d_w^{(0)}}.
\end{equation}
Knowing that the number $N(t)$ of jumps of the walker after an observation time $t$ scales as \cite{Bouchaud:1990b} 
\begin{equation}
\label{scalingN}
N(t)\sim t^\beta,
\end{equation}
the walk dimension $d_w$ of $X(t)$ can be written as
\begin{equation}
d_w=\frac{d_w^{(0)}}{\beta}.
\end{equation}

The process $X_n$ has by definition stationary increments and we thus have from section \ref{arguetheta}
\begin{equation}
\theta^{(0)}=1-\frac{d_f}{d_w^{(0)}}.
\end{equation}



Last, the persistence exponent $\theta$ is  known \cite{Bray:2013} to be related to $\theta^{(0)}$ by
\begin{equation}
\theta=\beta \theta^{(0)},
\end{equation}
as is  found by using the definitions of the persistence exponents $\theta$ and $\theta^{(0)}$, and again the scaling of $N(t)$ given by Eq. (\ref{scalingN}).
With these results, we finally obtain in the compact case that 
\begin{equation}
T_{\rm typ}=R^{d_w^{(0)}/\beta}
\end{equation}
and
\begin{equation}
G(\eta;a,r,R)=h(\eta)\left(\frac{r}{R}\right)^{d_w^{(0)}\theta^{(0)}}=h(\eta)\left(\frac{r}{R}\right)^{d_w^{(0)}-d_f}.
\end{equation}
Note that the scalings on the geometrical parameters $r$ and $R$ of \cite{Condamin:2007yg,Meyer:2011} are  recovered as specific cases of our general results. 

In the non-compact case, 
the distribution of $\tau_n$ has a power law tail:
\begin{equation}
P(\tau_n)\sim \frac{{\widetilde \tau}^\beta}{\tau_n^{1+\beta}}.
\end{equation}

We can directly apply Eq.~(\ref{generalisation}) with $\gamma=\beta$ and $\alpha=0$:
\begin{equation}
T_\mathrm{typ}\sim R^{d_w}\left(\frac{R}{a}\right)^{\frac{\psi}{\beta}}.
\end{equation}
Here, because $\psi$ is a geometrical quantity, its value is the same as that of the discrete walk $X_n$:
\begin{equation}
\psi=d_f-d_w^{(0)}=d_f-d_w \beta.
\end{equation}
Note that the formula \eqref{formulepsi} does not directly apply, because of divergences of both numerator and denominator of \eqref{rapport}. 
Finally,
\begin{equation}
T_\mathrm{typ}\sim \frac{R^{d_f/\beta}}{a^{d_f/\beta-d_w}}
\end{equation}
and  the distribution of the rescaled FPT $\eta=T/T_\mathrm{typ}$ reads
\begin{equation}
G(\eta;a,r,R)=h(\eta)\left[1-C\left(\frac{a}{r}\right)^{d_f-\beta d_w}\right].
\end{equation}
and we thus recover the results given in Refs.~\cite{Condamin:2007yg,Meyer:2011} as specific cases of our general results. 




\subsection{ ``Heavy tailed Random Acceleration Process''}

The previous paragraph explained why usual CTRWs do not exhibit aging for the quantities we focus on, we now turn to an example that combines heavy-tailed effects and aging features. We consider a $d_f$-dimensional Random Acceleration Process in discrete time, $\mathbf{x}(t_{i})$, naturally defined by :
\begin{align}
\mathbf{x}(t_{i+1})=\mathbf{x}(t_{i})+\mathbf{v}(t_{i}) \\
\mathbf{v}(t_{i+1})=\mathbf{v}(t_{i})+\boldsymbol{\xi}(t_{i}) 
\label{HeavyTailedLangevin}
\end{align}
where $\boldsymbol{\xi}$ is a $d_f-$dimensional vector, whose coordinates are independent random variables of zero mean and unit variance. Here we assume that the duration of each step, $\tau_i\equiv t_{i+1}-t_i$, is drawn from a L\'evy law of heavy  tail parameter $\gamma$, \textit{i.e.} $\rho(\tau_i)\sim_{\tau_i\rightarrow\infty}1/\tau_i^{1+\gamma}$. This obtained process is called ``Heavy-tailed Random Acceleration Process'' and combines aging effects and heavy tails by construction.

Since the walk dimension of the standard RAP is $2/3$, the walk dimension of the heavy tailed RAP is $d_w=2/(3\gamma)$. Using the fact that the aging exponent is equal to one for the standard RAP, we obtain an aging exponent for the heavy tailed RAP $\alpha=\gamma$. Moreover, the transience exponent $\psi$ is not affected by the waiting times, so that we still have $\psi=d_f-1$. 
By applying the results of Eq.(\ref{generalisation}), we obtain for $d_f>1$
\begin{equation}
T_{\text{typ}}=R^{\frac{2}{3\gamma}}\left(\frac{R}{a}\right)^{\frac{2(d_f-1)}{3\gamma}}
\end{equation}


\textbf{Simulation details (Fig. 4H).} The above relation is checked by means of 2d numerical simulations with $\gamma=0.7$ that are carried out by using directly the discretized Langevin equation \ref{HeavyTailedLangevin} . At each time step, we add a normal variable to the velocity, and we draw a waiting time $\tau$ from a $\gamma$-stable distribution, and the position of the walker is actualized. Simulations are done in 2$d$, in a confining volume $V=R^2$ with periodic boundary conditions. The initial distance between the walker and the target is $r=10$ and the radius of the target is $a=1$.

\end{document}